\documentclass[12pt,preprint]{aastex}
\usepackage[pdftex]{}

\newcommand  \kms      {\ifmmode {\rm km\,s}^{-1} \else km\,s$^{-1}$\fi}
\newcommand  \cc       {\hbox{cm$^{-3}$}}
\newcommand  \cmii     {\hbox{cm$^{-2}$}}
\newcommand  \ergs     {\ifmmode {\rm erg\,s}^{-1} \else erg s$^{-1}$\fi}
\newcommand  \ergcms   {\ifmmode {\rm erg\,cm}^{-2}\,{\rm s}^{-1}
                        \else erg\,cm$^{-2}$\,s$^{-1}$\fi}
\newcommand  \ergcmsA {\ifmmode{\rm erg\,cm}^{-2}\,{\rm s}^{-1}\,{\rm\AA}^{-1}
                        \else ergs\,cm$^{-2}$\,s$^{-1}$\,\AA$^{-1}$\fi}
\newcommand \ergcmsHz {\ifmmode{\rm erg\,cm}^{-2}\,{\rm s}^{-1}\,{\rm Hz}^{-1}
                        \else ergs\,cm$^{-2}$\,s$^{-1}$\,Hz$^{-1}$\fi}
\newcommand  \phcms    {\ifmmode {\rm ph\,cm}^{-2}\,{\rm s}^{-1}
                        \else ,ph\,cm$^{-2}$\,s$^{-1}$\fi}
\newcommand  \phcmsA   {\ifmmode {\rm ph\,cm}^{-2}\,{\rm s}^{-1}\,{\rm\AA}^{-1}
                        \else ph\,cm$^{-2}$\,s$^{-1}$\,\AA$^{-1}$\fi}
\def\micron{\ifmmode \mu{\rm m} \else $\mu$m\fi}
\def\kms{\ifmmode {\rm km\,s}^{-1} \else km\,s$^{-1}$\fi}
\def\Hubble{\ifmmode {\rm km\,s}^{-1}\,{\rm Mpc}^{-1}
        \else km\,s$^{-1}$\,Mpc$^{-1}$\fi}
\def\ergsec{\ifmmode {\rm ergs\;s}^{-1} \else ergs s$^{-1}$\fi}
\def\ergscm{\ifmmode {\rm ergs\,s}^{-1}\,{\rm cm}^{-2}
          \else ergs\,s$^{-1}$\,cm$^{-2}$\fi}
\def\ergscmA{\ifmmode {\rm ergs\,s}^{-1}\,{\rm cm}^{-2}\,{\rm \AA}^{-1}
          \else ergs\,s$^{-1}$\,cm$^{-2}$\,\AA$^{-1}$\fi}
\def\ergscmHz{\ifmmode {\rm ergs\,s}^{-1}\file:///home/schweitzer/tex/nH21_5/silicate15august.ps,{\rm cm}^{-2}\,{\rm Hz}^{-1}
q          \else ergs\,s$^{-1}$\,cm$^{-2}$\,Hz$^{-1}$\fi}
\def\Msun{\ifmmode M_{\odot} \else $M_{\odot}$\fi}
\def\Lsun{\ifmmode L_{\odot} \else $L_{\odot}$\fi}
\def\qo{\ifmmode q_{0} \else $q_{0}$\fi}
\def\Ho{\ifmmode H_{0} \else $H_{0}$\fi}
\def\ho{\ifmmode h_{0} \else $h_{0}$\fi}
\def\qo{\ifmmode q_{0} \else $q_{0}$\fi}
\def\ao{\ifmmode a_{0} \else $a_{0}$\fi}
\def\to{\ifmmode t_{0} \else $t_{0}$\fi}
\def\Halpha{\ifmmode {\rm H}\alpha \else H$\alpha$\fi}
\def\Hbeta{\ifmmode {\rm H}\beta \else H$\beta$\fi}
\def\hb{\ifmmode {\rm H}\beta \else H$\beta$\fi}
\def\Hgamma{\ifmmode {\rm H}\gamma \else H$\gamma$\fi}
\def\Hdelta{\ifmmode {\rm H}\delta \else H$\delta$\fi}
\def\Lya{\ifmmode {\rm Ly}\alpha \else Ly$\alpha$\fi}
\def\Lyb{\ifmmode {\rm Ly}\beta \else Ly$\beta$\fi}
\def\hi{\ifmmode \mbox{{\rm H}\,{\sc i}} \else H\,{\sc i}\fi}

\def\ciii{\ifmmode {\rm C}\,{\sc iii} \else C\,{\sc iii}\fi}

\def\o5007{[O\,{\sc iii}]\,$\lambda5007$}

\def\ne212m {[Ne\,{\sc ii}]\,$12.8 \mu$m}
\def\nulnu60 {$\nu$L$_\nu$(60$\mu$m)}

\def  \kms         {\hbox{km s$^{-1}$}}          
\def  \ergs        {\hbox{erg s$^{-1}$}}              
\def  \cc          {\hbox{cm$^{-3}$}}

\def  \cmii        {\hbox{cm$^{-2}$}}

\def  \La          {\ifmmode {\rm Ly}\alpha \else Ly$\alpha$\fi}
\def  \Ka          {\ifmmode {\rm K}\alpha \else K$\alpha$\fi}
\def  \Lb          {\ifmmode {\rm L}\beta \else L$\beta$\fi}
\def  \Ha          {\ifmmode {\rm H}\alpha \else H$\alpha$\fi}
\def  \Hb          {\ifmmode {\rm H}\beta \else H$\beta$\fi}
\def  \Pa          {\ifmmode {\rm P}\alpha \else P$\alpha$\fi}
\def  \CIIIb       {\ifmmode {\rm C}\,{\sc iii]}\,\lambda1909
                     \else C\,{\sc iii]}\,$\lambda1909$\fi}
\def  \CIV         {\ifmmode {\rm C}\,{\sc iv}\,\lambda1549
                     \else C\,{\sc iv}\,$\lambda1549$\fi}
\def  \MgII         {\ifmmode {\rm Mg}\,{\sc ii}\,\lambda2798
                     \else Mg\,{\sc ii}\,$\lambda2798$\fi}
\def  \OVI         {\ifmmode {\rm O}\,{\sc vi}\,\lambda1035
x
                     \else O\,{\sc vi}\,$\lambda1035$\fi}

\def \spitzer      {{\it Spitzer}}

\def \IRS      {{\it IRS}}

\newcommand{\mapiii}{MAPPINGS  \textsc{iii}}

\def  \kms         {\hbox{km s$^{-1}$}}          
\def  \ergs        {\hbox{erg s$^{-1}$}}              
\def  \cc          {\hbox{cm$^{-3}$}}

\def  \cmii        {\hbox{cm$^{-2}$}}

\def  \La          {\ifmmode {\rm Ly}\alpha \else Ly$\alpha$\fi}
\def  \Ka          {\ifmmode {\rm K}\alpha \else K$\alpha$\fi}
\def  \Lb          {\ifmmode {\rm L}\beta \else L$\beta$\fi}
\def  \Ha          {\ifmmode {\rm H}\alpha \else H$\alpha$\fi}
\def  \Hb          {\ifmmode {\rm H}\beta \else H$\beta$\fi}
\def  \Pa          {\ifmmode {\rm P}\alpha \else P$\alpha$\fi}
\def  \CIIIb       {\ifmmode {\rm C}\,{\sc iii]}\,\lambda1909
                     \else C\,{\sc iii]}\,$\lambda1909$\fi}
\def  \CIV         {\ifmmode {\rm C}\,{\sc iv}\,\lambda1549
                     \else C\,{\sc iv}\,$\lambda1549$\fi}
\def  \MgII         {\ifmmode {\rm Mg}\,{\sc ii}\,\lambda2798
                     \else Mg\,{\sc ii}\,$\lambda2798$\fi}
\def  \OVI         {\ifmmode {\rm O}\,{\sc vi}\,\lambda1035
x
                     \else O\,{\sc vi}\,$\lambda1035$\fi}

\def \spitzer      {{\it Spitzer}}

\def \IRS      {{\it IRS}}

\shorttitle{Extended silicate dust emission in PG QSOs}
\shortauthors{Schweitzer et al.}

\begin{document}

\title{Extended silicate dust emission in PG QSOs}

\author{M. Schweitzer}
\affil{Max-Planck-Institut f\"ur extraterrestrische Physik\\
       Postfach 1312, 85741 Garching, Germany}
\email{schweitzer@mpe.mpg.de}

\author{B. Groves}
\affil{Sterrewacht Leiden, Niels Bohr Weg 2, 2333 Leiden, Netherlands}
\email{brent@strw.leidenuniv.nl}

 \author{H. Netzer}
\affil{School of Physics and Astronomy and the Wise Observatory, The Raymond and Beverly Sackler Faculty of Exact Science, Tel-Aviv University, Tel-Aviv 69978, Israel}
\email{netzer@wise.tau.ac.il}

\author{D. Lutz, E. Sturm, A. Contursi, R. Genzel, L.J. Tacconi}
\affil{Max-Planck-Institut f\"ur extraterrestrische Physik\\
       Postfach 1312, 85741 Garching, Germany}
\email{lutz@mpe.mpg.de, sturm@mpe.mpg.de, contursi@mpe.mpg.de, genzel@mpe.mpg.de, linda@mpe.mpg.de }

\author{S. Veilleux, D.-C. Kim, D. Rupke}
\affil{Department of Astronomy, University of Maryland, College Park, MD 20742-2421, USA}
\email{veilleux@astro.umd.edu, dckim@astro.umd.edu, drupke@astro.umd.edu}

\author{A.J. Baker}
\affil{Department of Physics and Astronomy, Rutgers, the State University of New Jersey, 136 Frelinghuysen Road, Piscataway, NJ 08854-8019}
\email{ajbaker@physics.rutgers.edu}

\begin{abstract}
This paper addresses the origin of the silicate emission observed in PG  
QSOs, based on observations with the {\it Spitzer Space Telescope}.
Scenarios based on the unified model suggest that silicate emission in AGN arises mainly from the 
illuminated faces of the clouds in the torus at temperatures near sublimation. 
However, detections of silicate emission in Type 2 QSOs,
and the estimated cool dust temperatures, argue for a more extended emission region.
To investigate this issue we present the mid-infrared spectra of 23 QSOs. These spectra,
and especially the silicate emission features at $\sim 10$ and $\sim 18$ $\mu$m,
can be fitted using dusty narrow line region (NLR) models and
a combination of black bodies. The bolometric luminosities of the
QSOs allow us to derive the radial distances and covering factors for the silicate-emitting 
dust. The inferred radii are 100-200 times larger than the dust sublimation radius,
much larger than the expected dimensions of the inner torus. Our QSO mid-IR spectra
are consistent with the bulk of the silicate dust emission arising from the dust in the innermost parts of the NLR.
\end{abstract}


\section{Introduction}

Unified schemes for active galactic nuclei (AGN) 
postulate an obscuring torus surrounding an accreting super-massive black 
hole. Models predict that the infrared spectral energy distribution (SED) 
of the torus depends sensitively on its orientation, geometry and density 
distribution \citep[e.g.][]{pier92,granato94,efstathiou95,granato97,nenkova}. 
In particular, the tori are predicted
to exhibit prominent silicate dust features in either absorption or emission,
depending on whether an AGN is viewed with the torus edge-on (Type 2) or
face-on (Type 1). Previous failures to detect strong 9.7$\mu$m silicate 
emission in Type 1 AGN led to several proposed modifications of the unified 
model. For example modified grain size distributions have been assumed 
\citep{laordrain1993,maiolino} or a clumpiness of the torus invoked 
\citep{nenkova}.

The {\it Spitzer Space Telescope} ({\it Spitzer}), with its good mid-infrared (mid-IR) wavelength coverage and sensitivity,
has drastically changed our view of this problem. 
\citet{siebenmorgen05a} and \citet{hao05} reported the
first \spitzer\ Infrared Spectrograph (IRS) detections of prominent silicate emission features in the
mid-infrared spectra of several luminous quasars. Sturm et al. (2005) 
reported the first detection of 10 and 18 $\mu$m silicate emission features in a low-luminosity LINER (NGC 3998).
Comparison to the 10/18 $\mu$m feature ratio of optically thin emission from silicate dust at different temperatures suggests a modest temperature ($\sim$200K) of the emitting dust. 

The presence of prominent 10$\mu$m silicate emission features in AGN covering a broad range in 
luminosity could be taken as direct evidence for the existence of an
obscuring torus. However, it is not at all clear whether this emission actually arises
from the inner regions of a face-on torus. 
Depending on the size and composition of the grains, sublimation
occurs between $\sim$~800 and 1500 K \citep{kimura} which is also the temperature range expected 
for silicate dust located near the hot inner torus wall. 

The lower temperature indicated for the emitting silicate dust can be interpreted as evidence for dust emission from regions located further away from the central heating source. Several arguments support such a scenario. Silicate emission has also been detected in Type-2 QSOs \citep{sturm2006,teplitz06},
whereas for an edge-on view of the torus, one would expect to see silicate 
in absorption only. An extended emitting region, with dimensions much larger than the inner torus dimension, is fully consistent with this result. Broad-band $10\mu m$ imaging of several nearby AGN suggests extended
mid-infrared continuum \citep{cameron93,tomono,bock00,radomski03,packham05}.
\citet{efstathiou06} has modeled the silicate emission of the Type-2 QSO
IRASF10214+4724 \citep{teplitz06} invoking extended NLR dust in addition to an AGN torus.
\citet{marshall07} conclude that some of the optically thin warm emission in the QSO PG0804+761 may emerge from regions beyond the torus and suggest clouds in the NLR as a possible origin of this emission. 
These arguments indicate that silicate emission may originate in extended
regions ($\sim 100$ pc dimension).

The nature and location of the extended silicate-emitting region is not yet known. In this paper we explore
in a quantitative way one plausible interpretation, namely the association of this cool dust with the NLR.
To this end, we present fits of our QSO spectra with a superposition of NLR dust models and spectral components 
representing the innermost hot dust and the bulk of the inner structure emission (both related to the torus) as well as the large scale host emission. The fitted model and the bolometric luminosity of each source enable us to estimate the cool dust distance and its covering factor. We note that our models cannot exclude the possibility of a torus contribution to the observed silicate emission.

In section \ref{sec:Obs} we describe our QSO sample. In section \ref{sec:model}, we introduce the model components
and detail our fitting procedure. We also
describe how we estimate the silicate dust cloud distances and the related covering
factors. In section \ref{sec:robustness}, we discuss the dependence on model parameters. In section \ref{sec:results}, we present the results of the fits, which are then discussed in section \ref{sec:discussion}. Finally section
\ref{sec:conclusions} summarizes our conclusions.

\section{}
\subsection{The PG QSO Sample}\label{sec:Obs}

The sources used in our study are part of the \spitzer\ spectroscopy component (PID 3187, PI Veilleux) of the QSO/ULIRG evolutionary study (QUEST). The project and the sample are described in \citet{schweitzer06} (hereafter paper I). The QSO sample is largely drawn from that of 
\citet{guyon02} and \citet{guyon06}. It consists primarily of
Palomar-Green (PG) QSOs \citep{schmidt83} and covers the full ranges of bolometric luminosity $\sim 10^{11.5-13}$\Lsun \citep[based on the absolute $B$ band magnitude and the SED of][]{elvis94}, radio loudness, and infrared excess (\nulnu60 /L$_{Bol}$ $\sim$ 0.02--0.35) 
spanned by the local members of the PG QSO sample  \citep[see also][for a 
recent view on selection effects in the PG sample]{jester05}. 
B2 2201+31A is not a PG QSO but is included in the sample because its $B$ magnitude actually satisfies the PG QSO 
completeness criterion of \citet{schmidt83}. The QUEST sample used in this paper includes 23 of 32 objects from the Guyon sample.
We add one Palomar-Green object from the Guyon et al. sample previously observed by \spitzer\ (PG0050+124 = IZw1; \citet{weedman05}). Table~\ref{tab:targets} lists names and redshifts of 
all 23 QSOs in our sample, six of which are radio-loud. This sample covers a range from $M_B= -21$ to $M_B=-26$, with median $M_B=-23.3$. We assume a cosmology with H$_0$=70km\,s$^{-1}$\,Mpc$^{-1}$, $\Omega_m=0.3$ and $\Omega_{\Lambda}=0.7$ throughout the paper.

\subsection{Data reduction}

For the QSO sample, spectra were taken both at 5-14$\mu$m in the low-resolution 
(SL short-low) mode and at 10-37$\mu$m in the high-resolution 
(SH short-high and LH long-high) modes of the \IRS\ (Houck et al. 2004). Slit widths of 
3.6\arcsec\ to 11.1\arcsec\ include much of the QSO hosts as well as the vicinity of the AGN. 
Our data reduction starts with the two-dimensional basic calibrated data
(BCD) products provided by version 12 of the \spitzer\ pipeline
reduction. We used our own IDL-based tools for removing outlying values 
for individual pixels and for sky subtraction,
and SMART \citep{higdon04} for extraction of the final spectra. Small 
multiplicative corrections were applied to stitch together
the individual orders of the low-resolution and high-resolution spectra, as well as additive corrections
for residual offsets still found between the low-resolution spectra
and the SH and LH high-resolution spectra after zodiacal light correction 
of the latter. Paper I discusses in greater detail the {\it Spitzer} IRS observations and our data reduction procedure.

\begin{deluxetable}{lllcc}
\tabletypesize{\footnotesize}
\tablewidth{0pt}
\tablecaption{QSO sample\label{tab:targets}}
\tablehead{
\colhead{Object}&
\colhead{z}&
\colhead{D$_L$}&
\colhead{log ($L_{5100}/erg s^{-1})$}&
\colhead{radio loud (L)/quiet (Q)}\\
\colhead{}&
\colhead{}&
\colhead{Mpc}&
\colhead{}&
\colhead{}\\
\colhead{(1)}&
\colhead{(2)}&
\colhead{(3)}&
\colhead{(4)}&
\colhead{(5)}
}
\startdata
PG0026+129          &0.1420 &672&  44.66 &Q\\
PG0050+124 (IZw1)   &0.0611 &274&  44.30 &Q\\ 
PG0838+770          &0.1310 &615&  44.16 &Q\\
PG0953+414          &0.2341 &1170 &45.11 &Q\\
PG1001+054          &0.1605 &768&  44.25 &Q\\
PG1004+130          &0.2400 &1203& 45.23 &L\\
PG1116+215          &0.1765 &853&  45.13 &Q\\
PG1126-041 (Mrk1298)&0.0600 &269&  43.82 &Q\\
PG1229+204 (Mrk771) &0.0630 &283&  44.13 &Q\\
PG1302-102          &0.2784 &1425& 45.17 &L\\
PG1309+355          &0.1840 &893&  44.81 &L\\
PG1411+442          &0.0896 &410&  44.31 &Q\\
PG1426+015          &0.0865 &395&  44.44 &Q\\
PG1435-067          &0.1260 &590&  44.39 &Q\\
PG1440+356 (Mrk478) &0.0791 &359&  44.22 &Q\\
PG1613+658 (Mrk876) &0.1290 &605&  44.70 &Q\\
PG1617+175          &0.1124 &522&  44.29 &Q\\
PG1626+554          &0.1330 &626&  44.44 &Q\\
PG1700+518          &0.2920 &1505& 45.68 &Q\\
PG2214+139 (Mrk304) &0.0658 &296&  44.40 &Q\\
B2 2201+31A         &0.2950 &1553& 45.91 &L\\
PG2251+113          &0.3255 &1706& 45.63 &L\\
PG2349-014          &0.1740 &840&  45.21 &L\\
\enddata
\tablecomments{\\
Col. (1) --- Source name.\\
Col. (2) --- Redshift.\\
Col. (3) --- Luminosity distance in Mpc for a H$_0$=70km\,s$^{-1}$\,Mpc$^{-1}$, $\Omega_m=0.3$ and $\Omega_{\Lambda}=0.7$ cosmology\\
Col. (4) --- Continuum luminosity $\lambda L_{\lambda}$ at $5100$\AA\ rest wavelength (from spectra by T. Boroson, taken from \citet{netzer07})\\
Col. (5) --- radio loudness for PG QSOs taken from Sanders (1989) and for B2 2201+31A from Hutchings \& Neff (1992)
}
\end{deluxetable}

\section{Modeling the PG QSO IRS Spectra}
\label{sec:model}

\subsection{Model Components}
\label{sec:modelcomp}

We have developed a procedure to fit  the spectra of our sources with different components that account for the presence of
optically thick AGN-heated dust emission, silicate dust emission and reprocessed stellar emission from the dust in star-forming regions of the host galaxy.

Following recent AGN models \citep[e.g.][]{siebenmorgen05b,
Hoenig06,Elitzur06}, we assume a hot dust component 
located near the central accretion disc that is mainly responsible for the
shortest wavelength infrared continuum emission. AGN models typically consider
this dust to be located within a torus-like structure. The sublimation temperature for
silicate dust grains ranges from 800 K to 1500 K \citep{kimura}. This is also the
temperature range expected for dust located at the inner surface of the
torus with a typical sublimation radius of roughly $R_{sub}\simeq0.5\cdot\sqrt{L_{bol46}}$ pc, where the bolometric luminosity, $L_{bol46} $, is given in units of $10^{46}$ erg/s \citep[e.g.,][with dependence on grain material and size]{barvainis87, barvainis92, granato97, nenkova}. 
To account for this component, we introduce a hot black body spectrum with a temperature restricted to range between 1000 K and 1700 K. To take the temperature distribution within the torus into account as well, we introduce two additional black bodies as fit components (section \ref{subsec:fitproc}) with temperatures between 150 K and 1000 K. In the PG QSOs, these three components represent the bulk of the torus emission. The black body temperatures are free parameters allowed to vary continuously between their specified limits. 

The fourth component represents cool dust. As explained in paper I and in \citet{netzer07} (hereafter paper II), this component is assumed to be dominated by reprocessed stellar emission from star forming regions in the host galaxy. To
account for this component we introduce a cool black body with temperature limited to 35-65 K, a typical range in starburst galaxies. In paper I we demonstrated that on average the 7.7PAH/FIR ratio in our PG QSOs is the same as in starburst dominated, local ULIRGs. This fact indicates similar properties for the star formation in both galaxy types. Hence the assumption of a cool black body (T=35-65 K) to account for the reprocessed stellar emission from star forming regions within the host galaxy is reasonable. We note that the IRS spectra do not include a substantial part of the FIR emission of the QSOs. This limitation tends to shift the inferred best-fit temperature of this component towards higher values. In addition to intrinsic variations of the temperature of starburst dust, this bias is the reason we introduce a limited temperature range for this component. 

To account for the contribution of star forming regions to the 4-40 $\mu$m spectra of our sources,
we use the ISO-SWS mid-IR spectrum of M82 \citep[from][]{sturm00} as a starburst
template spectrum. In particular, this template is used to fit the
broad polycyclic aromatic hydrocarbon (PAH) emission features found in
the {\it Spitzer}-IRS range, which are
clearly seen in some of our objects. These features are typical
indicators for recent star formation.
The ratio of PAH emission to FIR emission in galaxies is related to the average radiation field intensity, echoed in the 60/100 $\mu$m flux ratio tracing the large grain temperature \citep{lu03}.
By scaling the contribution of the cool black body relative to the contribution of the M82 template, the fitting routine has the freedom to take such a difference between M82 and the PG QSOs into account.
Using starburst-dominated ULIRG templates instead of M82 would be an alternative possibility. To exclude ambiguities regarding the power source (AGN or starburst) of the template spectrum, we prefer here to use M82 which, is a well studied, typical starburst galaxy.
We have also tested the fits using the starburst template of Brandl et al. (2006).
We found no significant differences (in $\psi^2$ or the contributing components) compared to the fits using the M82 template. This is due to the overall small contribution of the starburst template and the similarity between M82 and the Brandl template.
The contribution of the M82 starburst template is found to be consistent with the PAH detections in paper I.
 
An important new ingredient of the present work is dust emission from the NLR.
To fit the silicate emission, which is visible in nearly all of our QSO spectra, we use dust NLR models
based on those of \citet{groves}. In contrast to torus models, our NLR models account for cooler, optically thin dust that is located further away from the central AGN.
The code used to generate these models, \mapiii, encompasses all dust-related processes, including stochastic
heating, which allows individual small grains to reach high
temperatures when heated by energetic photons.  
For the fits to the IRS spectra, we have removed the line emission in order to concentrate on the IR continuum and the silicate features. 
The main physics of the AGN IR modeling, including the dust composition and incident AGN
spectrum, have been discussed in detail in \citet{groves}, and we briefly recapitulate the main parameters here.

\begin{figure}
\begin{center}
\includegraphics[width=250pt]{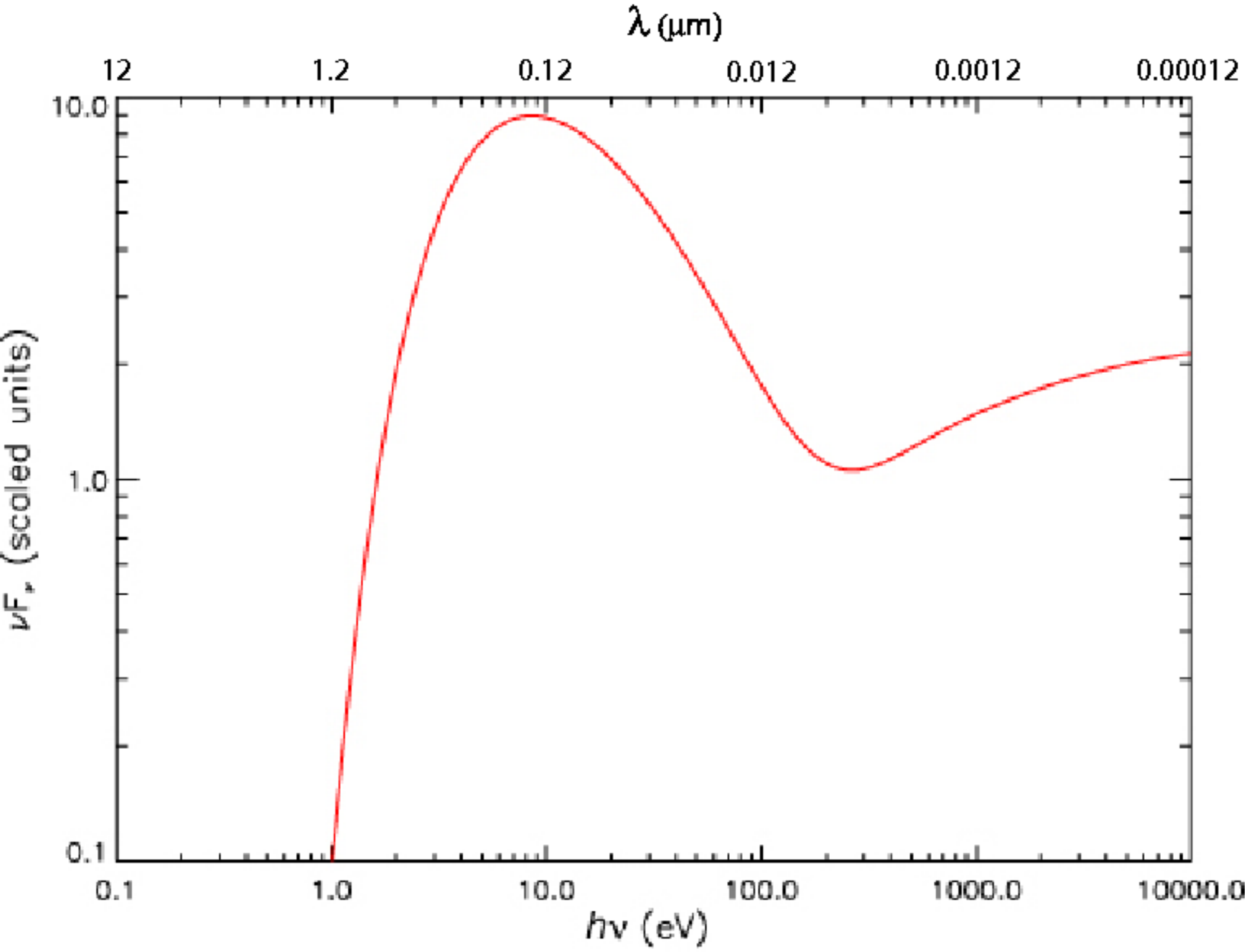}
\caption{Incident accretion disc spectrum from \citet{groves}.}
\label{fig:incident}
\end{center}
\end{figure}

In Figure \ref{fig:incident} we show the incident heating spectrum, which is a fit by \citet{groves} of two power-laws with exponential cut offs to the observations of \citet{elvis94}. We assume
the gas abundances have their solar values \citep{Asplund05}. Similarly,
the dust depletions are based on local measurements \citep{Kimura03} for these
models \citep[see Table 1 in ][]{groves}. We note that the actual
metallicities of the QSOs in our sample may be higher, and the actual dust depletions uncertain. 
The NLR models assume a mixture of silicaceous and graphitic dust
\citep{laordrain1993,draine84,weingartner01} with a grain size distribution
arising from a modified grain shattering profile, leading to
a smooth exponential cut-off in terms of the grain mass at both ends of the
distribution:
\begin{equation}
dN(a)/da = k  a^{-3.3}
\frac{e^{-(a/a_\mathrm{min})^{-3}}}{1+e^{(a/a_\mathrm{max})^3}},
\label{eqn:GSprof}
\end{equation}
with $k$ defined by the dust-to-gas ratio and $a$ being the grain size. The minimum and maximum grain sizes
are 0.01$\mu$m and 0.25$\mu$m respectively, for both grain types
considered. PAHs are assumed to be destroyed within the harsh NLR environment. 

In contrast to previous constant-pressure cloud models, we have assumed a constant density structure of
$n({\rm H})= 10^{4}$\cc$ $ for simplicity, and explored a range of 13 ionization
parameters\footnote{The ionization parameter is a dimensionless ratio relating the ionizing flux to the particle density: $U=\frac{F_{*}^{\rm ion}}{n_{\rm H}c}$. } (incident fluxes) ranging from $\log U=1$ to $-3$ in steps of 0.3 dex. These values are typical for NLRs, with $\log U=1.0$ being an extreme case, and $\log U=-3$ becoming too cool to
contribute any significant silicate emission. The 13 ionization parameters correspond to the 13 NLR models (see Table \ref{tab:NLRmodels}), that will be used later to fit the spectra.
The density assumed is on the moderate to high end for NLRs, which are actually likely to be
stratified in density \citep[e.g.,][]{Groves04}. However, the assumption of constant density 
allows us to define accurately the incident ionizing flux, the
dominant parameter controlling the dust temperature, and hence the emission. If
a lower density of $n({\rm H})= 10^{3}$\cc\  is assumed, the general shape of
the IR continuum appears the same for a given incident flux. However, the
ionization parameter for a given incident flux increases by a factor of 10,
resulting in stronger higher ionization lines. The assumed density has no
influence on the estimated distances of clouds from the AGN and should be considered as an
approximation. In section \ref{sec:discussion}, we will discuss the NLR
properties also taking line emission (e.g., [NeV]) into account. 
Using the line emission, we find that a density of $n({\rm H})= 10^{4}$\cc$ $ 
is an upper limit for the density of the line emitting region.

\begin{table}
\caption{AGN IR Model Parameters}\label{tab:NLRmodels}
\begin{tabular}{lll}
\hline
Model Number & $\log(U)$ & log($\frac{\mbox{Incident Flux}}{\mbox{\ergcms}}$)\\
&&\\
\hline
\hline
1 & \phs1.0 & 5.56 \\
2 & \phs0.6 & 5.16 \\
3 & \phs0.3 & 4.86 \\
4 & \phs0.0 & 4.56 \\
5 & $-$0.3 & 4.26 \\
6 & $-$0.6 & 3.96 \\
7 & $-$1.0 & 3.56 \\
8 & $-$1.3 & 3.26 \\
9 & $-$1.6 & 2.96 \\
10 & $-$2.0 & 2.56 \\
11 & $-$2.3 & 2.26 \\
12 & $-$2.6 & 1.96 \\
13 & $-$3.0 & 1.56 \\
\end{tabular}
\tablecomments{The NLR models assume a constant density of\\
$n({\rm H})= 10^{4}$\cc$ $ for simplicity.}
\end{table}

\begin{figure}
\includegraphics[width=400pt]{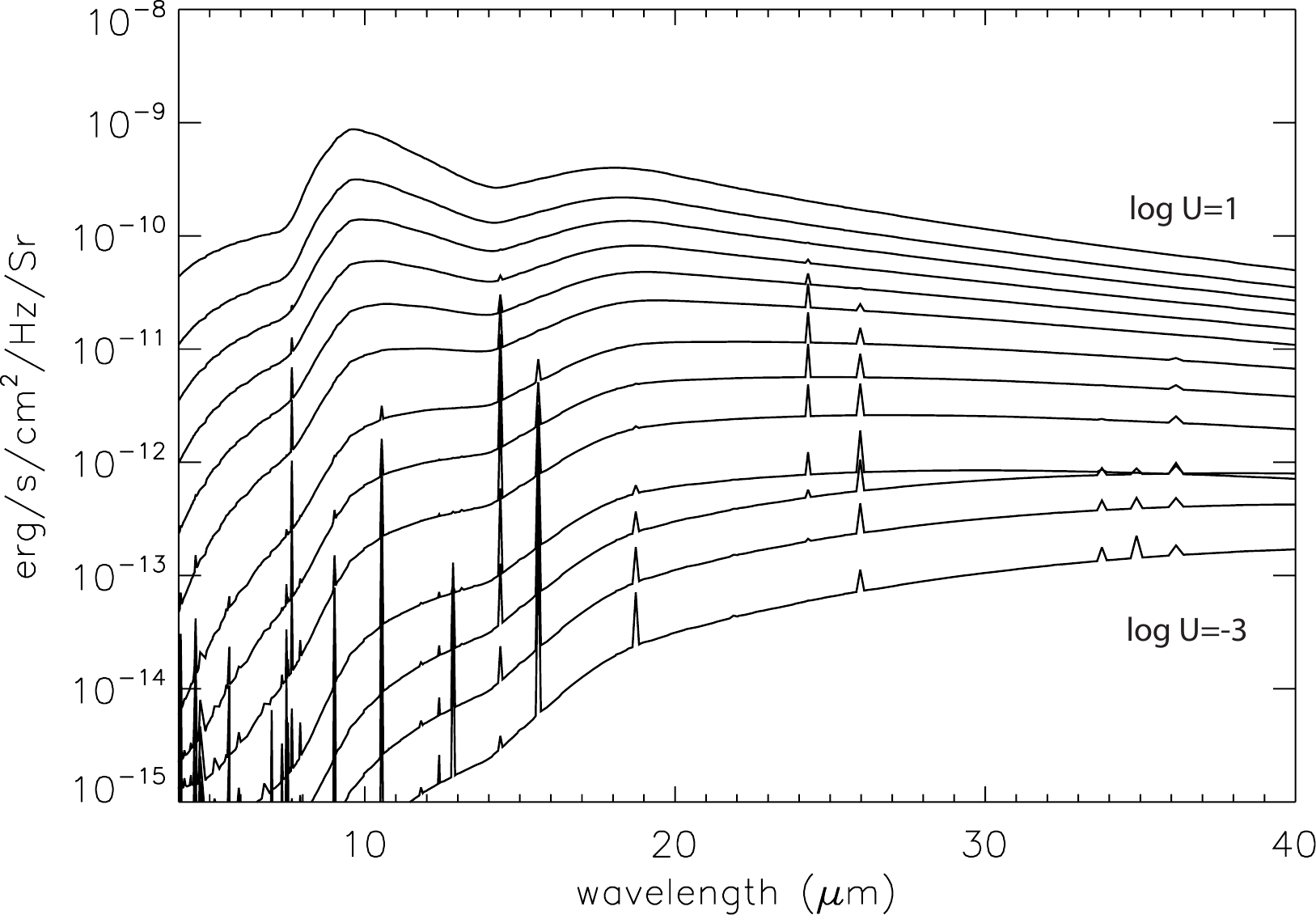}
\caption{Model NLR IR spectra from  $\log U=1.0$ (top model)  to $-3.0$ (bottom model) in steps of
~0.3 dex. }
\label{fig:modelIR}
\end{figure}
As in the \citet{groves} work, we have limited the column density of these models to $N($H$)=10^{21.5}$\cmii.
Fitting the QSO spectra with lower column densities (e.g., $N($H$)=10^{19}$\cmii) leads to unreasonably large covering factors.
Using a column density of $N($H$)=10^{21.5}$\cmii\ results in a median covering
factor of $\sim$0.16. Both the ratio of IR to optical/UV flux and {\it Hubble
Space Telescope} ({\it HST}) imaging of AGN \citep{kriss92} suggest a
$\sim30\%$ covering factor as an upper limit for the clouds in the NLR.
Increasing the column depth of the NLR models will only affect our results
minimally. By a column depth of $N(H)= 10^{21.5}cm^{-2}$, approximately 70 $\%$ of the heating flux has already been absorbed and re-emitted by gas and  dust, meaning that the preponderant shape of the IR emission is already in place. In addition, at larger column depths most of the dust is cool and emits most of its IR flux at wavelengths greater than 30$\mu$m, outside the range of the \emph{IRS} spectra.

The resulting nebular IR spectra are presented in figure \ref{fig:modelIR}. They
show the range in possible silicate emission, as well as the relevant emission
lines possible under these conditions assuming the above gas density.
We note again that our fits did not include the NLR emission lines, but in section \ref{sec:discussion}
we will discuss the implications for the NLR from a comparison of the modeled and measured line fluxes.

\subsection{Fitting Procedure}
\label{subsec:fitproc}
To fit the QSO spectra with the components presented in the previous
section we developed a template fitting tool based on the IDL routine MPFIT written by
C.B. Markwardt\footnote{http://cow.physics.wisc.edu/$\sim$craigm/idl/idl.html}. This
routine uses the
Levenberg-Marquardt technique and we use it here to minimize a modified $\chi^2$-value (hereafter called $\psi^2$) defined by
\begin{equation}
\psi^2=\sum_{\lambda}(\frac{F_{\rm obs}(\lambda)-F_{\rm mod}(\lambda)}{Err(\lambda)})^2,
\end{equation}
 where $F_{\rm obs}(\lambda)$ is the monochromatic QSO flux and $F_{\rm mod}(\lambda)$ is the
model flux made of the rebinned M82 template spectrum, the four black body spectra
and the NLR dust model spectra. 

Thus

\begin{equation}
F_{\rm mod}(\lambda)=\sum_{i=1}^{N_{\rm B}} a_i
\frac{B_i(\lambda,T_i[K])}{\max(B_i)}+ 
\sum_{i=1}^{N_{\rm template}} b_i \frac{F^{\rm
model}_i(\lambda)}{\max(F^{\rm model}_i)}, 
\end{equation}

where $B_i$ are the Planck functions. All models and template spectra are binned to the IRS spectrum resolution for our sources. The first term in $F_{mod}(\lambda)$ sums over the four black bodies, while the 
second sum includes all possible templates (in our case M82 and one (or more) of the 13 NLR models).
Each component of the model spectrum is normalized to its maximum within
the IRS range and then scaled by a factor $a_i$ or $b_i$. The
normalization prevents large variations in these factors.
The IRS spectra of our PG QSOs have been fitted using the IRS rest wavelength ranges plotted in Figure 5-8.
These ranges vary from source to source depending on the intrinsic redshift of the object.

$Err(\lambda)$ usually represents the 1-sigma
uncertainty on $F_{\rm obs}(\lambda)$, but in our case
it is defined in a different way. From tests with different weighting schemes, we found that we best achieved a reasonable weighting and a good fit quality using: 
\begin{equation}
Err(\lambda)=\frac{F_{PL}(\lambda)}{\sqrt(\frac{\Delta \lambda}{\lambda})}
\end{equation}
The flux trend $F_{PL}$ is from a power law fit to the IRS source spectrum and does not trace individual features.
This choice has been made to achieve similar weights for the same relative flux deviations $(\frac{F_{\rm obs}(\lambda)-F_{\rm mod}(\lambda)}{F_{PL}})$ at different flux levels $F_{PL}$. $\Delta\lambda$ is the local wavelength sampling density in micrometers. To ensure that better sampled wavelength regions do not dominate the fit due merely to their finer sampling, we estimate $\Delta\lambda$ using a neighborhood of five wavelength bins at each wavelength, and we make $Err(\lambda)$ proportional to $\sqrt{\frac{1}{\Delta\lambda}}$.
The 5-37$\mu$m IRS spectra have spectral resolution varying between $\sim$60 at the short wavelength end and $\sim$600 at the long wavelength end.
Our weighting method ensures a good fit quality over the entire wavelength range while simultaneously being able to fit the PAH emission features located within lower flux regions.
To further increase the fit quality at shorter wavelength, $Err(\lambda)$ is proportional to $\sqrt{\lambda}$, which results in a stronger weighting at shorter wavelengths. 

As we focus only on the continuum emission, we have removed all emission lines from the original spectra and the NLR models. All lines were cut from the spectra, with the resulting gaps not considered by the fitting routine. We fit each source with one NLR model plus the four black bodies and the M82 starburst template and repeat this procedure for all NLR models.
To obtain the best fitting model template ($F_{\rm mod}$) for each of the NLR models, 
the contributions of the chosen NLR model and starburst template ($b_{i}$) and the black bodies ($a_i$) are
allowed to vary, as well as the black body temperatures ($T_i$) within
their specified limits (i.e., hot: 1700 K-1000 K, warm$_1$ and warm$_2$: 1000 K-150 K and cool: 65 K- 35 K).
Finally, for each of our PG QSOs, we determined the best fitting (minimum
$\psi^2$) model template out of all the NLR models.

We do not consider the black body spectra used in our fits to be physically meaningful. Instead, they are a reasonable, physically motivated approximation to a smooth underlying continuum, necessary to achieve (in combination with the NLR models) a good fit quality for our spectra. To account for the strong emission found at the shortest wavelengths within the IRS range, we have to introduce a hot continuum component (the hottest black body) for all of our QSO spectra. The limited MIR wavelength range we cover prevents a robust determination of all parameters of this component, since the bulk of its emission is not covered by the IRS range. The same is true for the coolest black body. To achieve a good fit quality, we additionally introduce two black bodies with intermediate temperatures. 
Using just one black body of intermediate temperature leads to degeneracies within the 10$\mu$m region, with some sources having the $\sim10\mu$m silicate feature reproduced by the black body instead of the NLR model. To prevent this, we introduce a second intermediate black body, which allows for a smoother underlying continuum emission. Including even more black bodies was not necessary since a satisfying level of fit quality is reached with only the four black bodies.
Since we do not interpret the black bodies as real physical components and have only a limited wavelength range, we will not discuss their properties.
Note that for some sources, the fitting routine automatically minimizes the contributions of certain fit components.
This effect can be most dramatically seen in PG 1626+554, shown in figure \ref{fig:fit results (A)}, for which only two components (one hot black body and one NLR model) contribute to the best fit model.

Beyond our standard fitting procedure, we investigated several alternatives. First, we tested the fits using grey bodies proportional to $\lambda^{-\alpha}$ with $\alpha=0,1,2$ instead of black bodies. This change had no significant effect on the fit results ($\psi^2$, contributing components) or their interpretation.
Next we tested for a non-thermal contribution to the mid-IR spectra of our sources 
by extrapolating a power-law model (based on radio data from NED) to the MIR.
We found that for the radio-quiet objects, the non-thermal contribution is a 
factor 1000-10000 weaker than the MIR emission at 10 $\mu$m and can be neglected. 
Three of the six radio-loud objects have an implied weak non-thermal contribution at 10 $\mu$m
at the level of 1-10\% of the observed continuum.
For the other three radio loud sources (PG1302-102, PG2251+113 and B2 2201+31A), the
non-thermal contribution could be significant. These sources have been fitted 
using a power law, with the two intermediate temperature black bodies removed.
We find that including a power law non-thermal component does 
not change our result, since it provides fairly smooth underlying emission similar to that provided by our black bodies. The identification of our sources as radio loud/quiet is 
listed in table \ref{tab:targets}. 

The fitting procedure also allows us to use multiple NLR models simultaneously. A comparison between use of one and multiple NLR models is discussed in section \ref{sec:numberofmodels}.
Additionally our fitting routine allows us to obscure single components by a foreground screen. The possible impact of extinction on our results is discussed in section \ref{sec:extinction}.

\subsection{Cloud Distances and Covering Factors}

To estimate the NLR cloud covering factors and distances (from the central source)
we need to use the intrinsic AGN bolometric luminosity.
In this context, we have to discriminate between primary radiation emitted by the accretion disk and secondary 
radiation like that reprocessed by the torus. For a detailed discussion of the bolometric correction, see paper II. 
We adopt here a bolometric correction defined by $L_{bol}=7\times L(5100)$, where $L(5100)$ is the $5100$\AA\ rest wavelength continuum luminosity $(\lambda L_{5100})$. As explained in paper II, this choice avoids a double counting problem, since it accounts only for primary radiation and does not include reprocessed dust emission.
A larger bolometric correction would imply larger cloud distances and smaller covering factors.
The values of $L(5100)$ adopted here are based on the Boroson \& Green (1992)
observations, which used a relatively small aperture. Given the high
luminosity AGN, the contributions of the host galaxies at this
wavelength are negligible. Since the ground-based spectroscopy does not show
any indication for a high extinction, the intrinsic $L(5100)$ and
the derived bolometric luminosities cannot be significant larger than the
values used in this work.
The values of $L(5100)$ for the 23 sources are listed in table \ref{tab:targets}.

Given the calculated integrated incident flux $F_{in}$ (see table \ref{tab:NLRmodels}) for the best $\psi^2$ model, we can obtain the dust cloud distance,

\begin{equation}
\label{eq:dist}
R_{\rm dust}=\sqrt{\frac{L_{\rm bol}}{4\pi F_{\rm in}}},
\end{equation}

This in turn allows us to calculate the required NLR covering factor $c$ given the NLR dust luminosity:

\begin{equation}
c=\frac{F_{NLR_{fit}}}{F_{R_{dust}}}\cdot\frac{D_L^2}{R_{dust}^2}
\end{equation}

where $F_{NLR_{fit}}$ is the NLR model flux estimated by integrating the fitted NLR model over the observed IRS wavelength range and $F_{R_{dust}}$ is the model flux, integrated over the rest-frame wavelength range, at the distance $R_{dust}$ from the central source. $D_L$ is the luminosity distance of the source (listed in table \ref{tab:targets}).
The distances and covering factors obtained in this way are listed in table \ref{tab:resultsa}.

\subsection{Fit components and uncertainties}
\label{sec:robustness}

In this section, we discuss the impacts of extinction and the use of multiple NLR components on the fits. We also demonstrate the fit quality by fitting different NLR models to the spectrum of PG1004+130.

\subsubsection{The Effect of Extinction}
\label{sec:extinction}
All of our Type 1 QSOs show blue continuum emission, indicating a low extinction $(A_V<2)$. Nevertheless, we tested the effect of a line of sight extinction on our fit results. For this we assumed a uniform screen with fixed extinction in the range $A_V=0-5$. This was applied to the hot and warm black bodies as well as to the NLR models using the extinction curve of \citet{draine03}. We then repeated the complete fitting procedure using the extinguished black body and model spectra. As expected, small $(A_V<2)$ extinction does not affect the fit results.
Using $A_V=2.5$ for some sources, the best fitting NLR model became slightly hotter.
For the largest $(A_V=5)$ case, we find that, on average, the best NLR model becomes slightly hotter, resulting in a cloud distance lower by a factor of $\sim2$. We assume this factor to be the upper limit on the uncertainty in the estimated distance due to extinction. All fits of the PG QSOs presented in the following sections assume $A_V=0$ to match the low observed extinction.
For comparison, we also tested the fits with the extinction curve of Chiar \& Tielens (2006).
In general, the fit quality became worse (larger $\psi^2$-values) compared to fits with the curve of \citet{draine03}, while the best fitting NLR model did not change significantly.
Due to the better fit quality, we use the extinction curve of \citet{draine03} in the following.

\subsubsection{The number of NLR components}
\label{sec:numberofmodels}
\begin{figure}
\includegraphics[width=400pt]{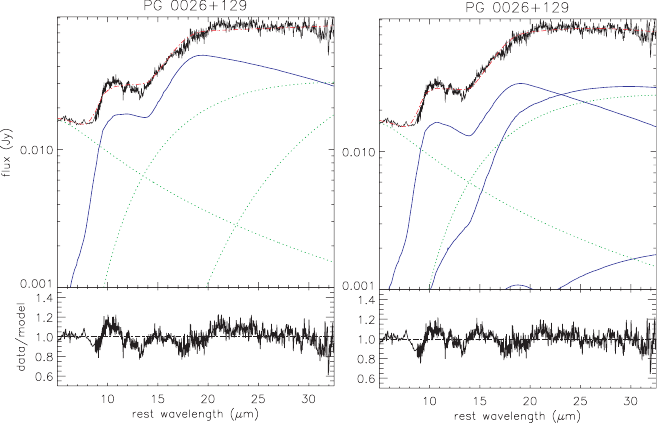}
\caption{Comparison of a single (left) and a multiple NLR model fit (right). In both fits similar cool NLR models produce the silicate emission features.( NLR model(s) (blue, solid curves); blackbodies (green, dotted curves); total model (red); observed spectrum (black) )}
\label{fig:multi}
\end{figure}

The fitting procedure described in section \ref{subsec:fitproc} assumes a single NLR component to infer the silicate dust distances and covering factors. Another possibility is contributions from several dusty clouds at different distances from the central source. To test how such a cloud distribution would influence our conclusions, we repeated the procedure using multiple NLR components. The fitting procedure then minimizes the contributions of those NLR components that do not increase the fit quality. The best fitting NLR models derived from this procedure have then been compared to the best fit single NLR model of the same source.
We found that for most sources, the best fit single NLR model is also the main contributor in the multiple NLR model fit, but this is sometimes accompanied by a second model (with similar $F_{in}$) but a smaller contribution.
Also, in some sources the cool black body is replaced by a cool NLR component whose temperature is too low to contribute to the silicate emission.
In three sources, the hottest and strongest NLR contributions come from components slightly hotter than those preferred by the single NLR fits. Figure \ref{fig:multi} shows a comparison between single and multiple NLR fits for PG0026+129.
The fit quality as judged by the value of $\psi^2$ does not change significantly (single model fit $\psi^2=0.0119$ ; multiple model fit $\psi^2=0.0097$).
In summary, our conclusion of extended silicate emitting dust is independent of the number of NLR components included in the fits. However, we cannot exclude the possibility of smaller contributions from other NLR components.

\subsubsection{Dependence on cloud distance}

\begin{figure}
\includegraphics[width=400pt]{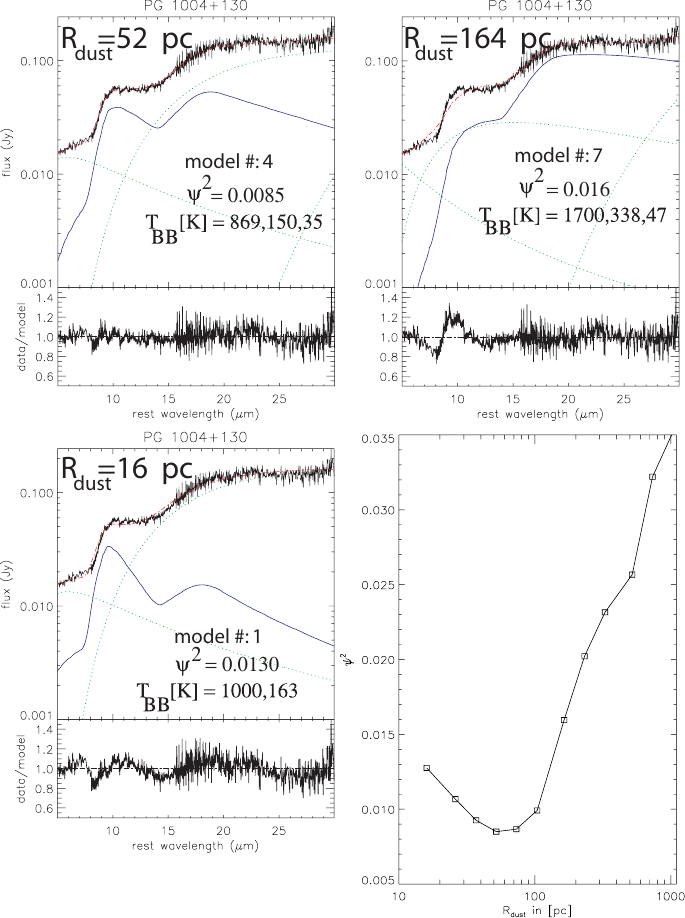}
\caption{Comparison of fit results for PG 1004+130 using the best fit model ($R_{dust}$=52 pc) or neighboring models differing by a multiplicative factor of $\sqrt{10}$ in radius
($R_{dust}$=16;164 pc). Conversion from NLR component to radius is via Eq.~5. For this example we also present the corresponding black body temperatures; however we caution that their physical meaning is limited due to the limited wavelength range we cover.}
\label{fig:bestfit}
\end{figure}

In fig. \ref{fig:bestfit} we demonstrate the quality of the fit for PG1004+130, which has been chosen as more or less representative for the whole sample. We compare the best-fit model with two fits using models that differ by a multiplicative factor of $\sqrt{10}$ in $R_{dust}$. Inspection of the residuals shows that these two fits are significantly worse than the best fit model. The mismatch can most clearly be seen at $\sim$10$\mu$m, where the neighboring models differ significantly from the data. In the lower right panel of fig. \ref{fig:bestfit} we show $\psi^2$ as a function of dust cloud distance for this QSO.
It can be seen that $\psi^2$ is a smooth function with a quite well defined minimum at a dust distance of $R_{dust}$=52 pc.

\section{Results}
\label{sec:results}
In figure \ref{fig:fit results (A)} we present the best fits for the complete
QSO sample. In the top half of each diagram, we show the best-fit
model with the observed spectrum. The data are shown in black
and the model in red. Also shown are the individual model components: the NLR models (blue, solid curve), 
the M82 starburst template with PAHs (blue, dashed curve) and the
four black bodies in green (dotted). The lower part of each figure shows the fit quality across the
spectrum. As can be seen, the fit is satisfactory (within 20$\%$ of the flux density) for all
sources. 

Table \ref{tab:resultsa} lists the best fitting NLR models for each
source in our sample and the estimated cloud distances and covering
factors. The cloud distances range from 9 pc (PG 0050+124) to 263 pc (PG 2251+113),
with a median value of 40 pc. This is larger than the expected dimension of the inner torus \citep[a few pc;][]{Jaffe04}.
These large radii are compatible with an extended silicate emission region. Since we do not use any torus model in our fits, we are not able to exclude the torus as the origin of the silicate emission. However, we have demonstrated that the NLR is a viable alternative as a source or contributor of the silicate emission. Using the average value of $F_{in}$ computed from our fits and Eqn.\ref{eq:dist}, the average scaling between radius and luminosity is given by:
\begin{equation}
 R_{dust}\simeq 80(L_{bol46})^{1/2} pc
\end{equation}
with the bolometric luminosity $L_{bol46} $ given in units of $10^{46}$ erg/s.
In figure \ref{fig:statistics}, we plot the distribution of the best fit models for the 23 sources. Each NLR model is related to a scaling relationship $R_{dust}\simeq x\cdot(L_{bol46})^{1/2}$pc, where $x$ is the scaling factor. 
The distribution peaks near $x=68$, which corresponds to model \# 5 with an incident flux of $10^{4.26} \frac{erg}{cm^2 s}$ (table \ref{tab:NLRmodels}). The model number (\#) is a running number (1-13) that identifies the NLR model. With increasing model number the NLR dust model becomes cooler and the scaling factor increases. In table \ref{tab:NLRmodels}, the model number is listed together with the ionization parameter and the incident flux assumed for the respective model. From figure \ref{fig:statistics}, it can also be seen that the hottest model ($x=15$) with the largest incident flux, is never the best fit model. The same holds true for models cooler than model \# 9. This indicates that the range of incident fluxes tried here is adequate to fit the dust properties in the NLRs of our 23 QSOs. The range in $R_{dust}$ defined by the hottest and coolest models is indicated by the dashed lines in figure \ref{fig:distance}.

The good fits we obtain support
the idea that the narrow line region is a possible candidate for hosting the relatively cool
silicate dust seen in emission.
In figure \ref{fig:distance}, we plot the estimated dust cloud distances versus the bolometric luminosities of the PG QSOs.
The solid line in this plot indicates the sublimation distance for the silicate grains calculated from $R_{sub}\simeq 0.5 (L_{bol46})^{1/2}$ pc (section \ref{sec:modelcomp}). Note that we assume here a direct exposure of the silicate grains by the central source. On average we find for the dust distance: 
\begin{equation}
R_{dust}\simeq 170 R_{sub}
\end{equation}
The dashed lines in figure \ref{fig:distance} indicate the distance range covered by our NLR dust models.
The typical NLR sizes in our sources are about 5-10 times larger than the largest distances we estimate for the NLR clouds.
We refer to section \ref{sec:lines} for a discussion of the NLR size.

We also compared our results to other known observations.
The lower triangle in figure \ref{fig:distance} is an upper limit on the torus radius estimated from MIR interferometric observations of the nucleus of the nearby Type 2 AGN, NGC 1068 \citep{Jaffe04}.
The upper triangle indicates the expected silicate dust distance as estimated from our average scaling relation for this source.
This example shows that the torus dimension is much smaller than what is expected for the NLR dust cloud distance in an AGN of this particular luminosity.
We will discuss this further in section \ref{subsec:distance}.
The crosses in figure \ref{fig:distance} compare our estimate using the average scaling relation (upper cross) for the silicate dust cloud distance in the Type 1 QSO PG 0804+761 with an estimate of \citet{marshall07} (lower cross) using a decomposition technique similar to ours. For this comparison, we assumed the bolometric AGN luminosity given in \citet{marshall07} of $L_{bol46}=0.48$. For PG 0804+761, the two estimates of $R_{dust}$ differ by a factor of 2.9 and are in good agreement considering the expected uncertainties and the general scatter of the cloud distances.

The cloud covering factors are listed in table \ref{tab:resultsa}. They range from 0.09
(B2 2201+31A) to 0.50 (PG1309+355), excluding PG1001+054 which is the only
source with a covering factor larger than one (1.16). This implausibly large value could
be the result of the source variation (e.g. S. Kaspi et al. (2000) , A. Wandel et al. (1999)) between the epochs of the
optical and the {\it Spitzer} IRS observations. The median covering factor for the full sample,
excluding PG 1001+054, is 0.16. In figure \ref{fig:covering}, we show the distribution of the covering factor excluding PG1001+054.

Table \ref{tab:resultsb} presents the contribution of
the NLR dust components to the total spectrum in two ways. In the second column, we compare the relative contribution to the 6-25$\mu$m rest-frame wavelength range. In the third column, we list the NLR contribution to the 15$\mu$m rest-wavelength continuum source flux. The table also shows the relative contribution of the M82 starburst template to the 6-25 $\mu$m rest-frame wavelength range. The upper limits for this contribution have been calculated by scaling the M82 template to match the 3$\sigma$ upper limits for the 7.7 PAH feature (paper I).
The contribution of the starburst template is strongly correlated with
PAH strength. Most objects which show clear PAH features in their spectra have a strong M82 contribution (see figure \ref{fig:fit
results (C)} for an example). 
The median contribution of the NLR dust to the 6-25 $\mu$m restframe emission is 24 $\%$ and the full range is from 15 to 51 $\%$.

In the literature, the strength of the silicate emission in AGN spectra has usually been expressed as an
equivalent width of the 9.7$\mu$m feature after interpolation of the underlying continuum emission.
However, caution is required in using this method, since the possibility of silicate emission hidden by the presence of star formation may bias those estimates. For example, in the fits of PG1440+356 and PG1613+658 in figure \ref{fig:fit results (C)}, 
it can be seen that this method would result in an underestimate of the silicate strength. 
The apparent smoothness of the spectrum around 10 $\mu$m is caused by the superposition of starformation related PAH features and a silicate dust contribution that is actually quite strong. 
To achieve a good estimate for the silicate strength in such composite sources, 
a decomposition procedure is desirable.

\begin{figure}
\begin{center}
\includegraphics[width=300pt]{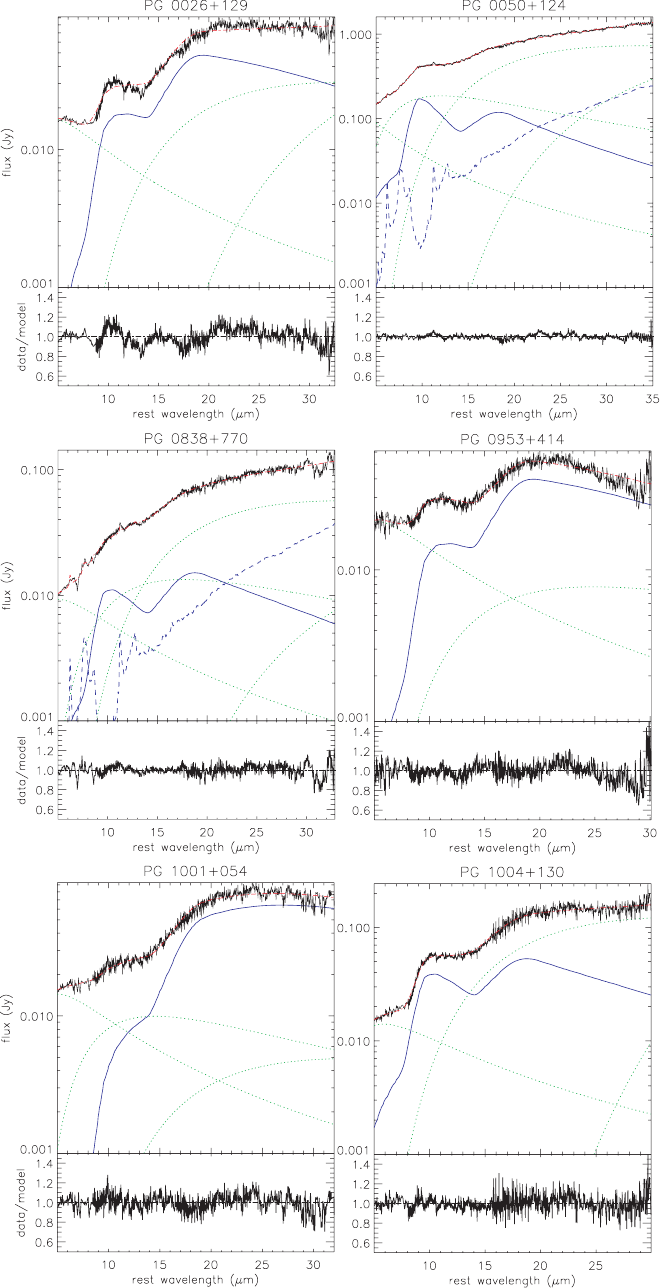}
\caption{fit results: NLR model (blue, solid curve); M82 (blue, dashed curve); blackbodies (green, dotted curves); total model (red curve); observed spectrum (black curve)}
\label{fig:fit results (A)}
\end{center}
\end{figure}

\begin{figure}
\begin{center}
\setcounter{figure}{4}
\includegraphics[width=300pt]{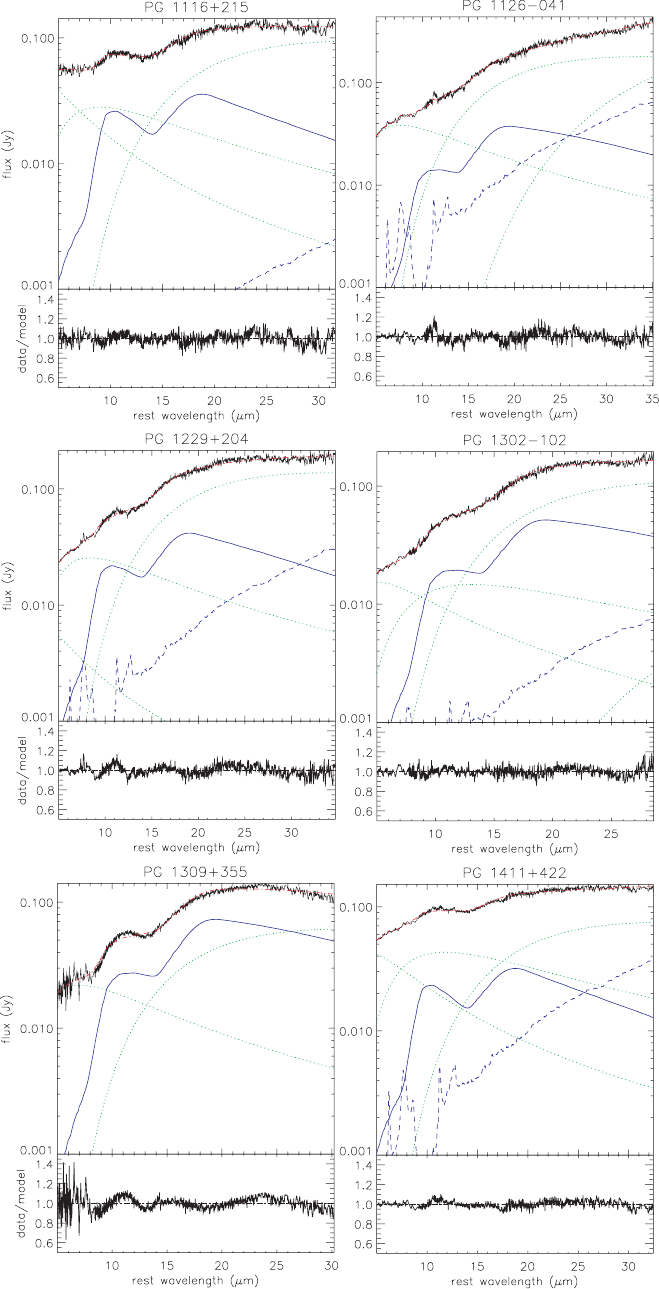}
\caption{continued}
\label{fig:fit results (B)}
\end{center}
\end{figure}

\begin{figure}
\begin{center}
\setcounter{figure}{4}
\includegraphics[width=300pt]{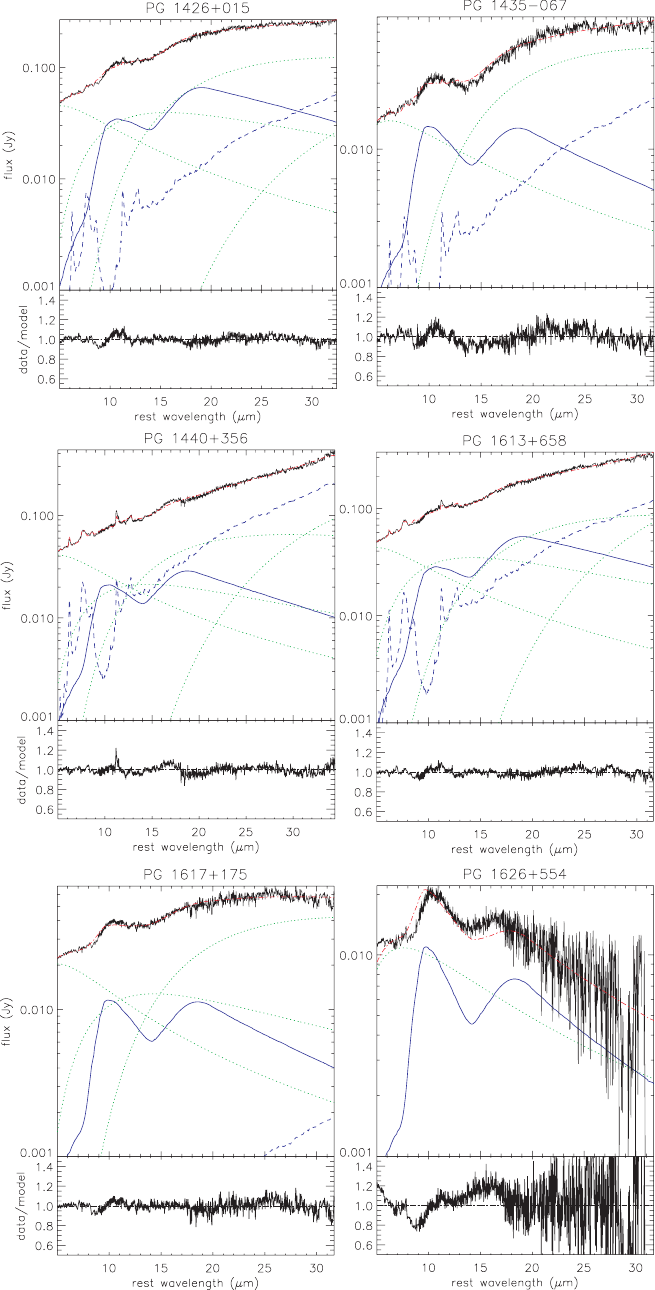}
\caption{continued}
\label{fig:fit results (C)}
\end{center}
\end{figure}

\begin{figure}
\begin{center}
\setcounter{figure}{4}
\includegraphics[width=300pt]{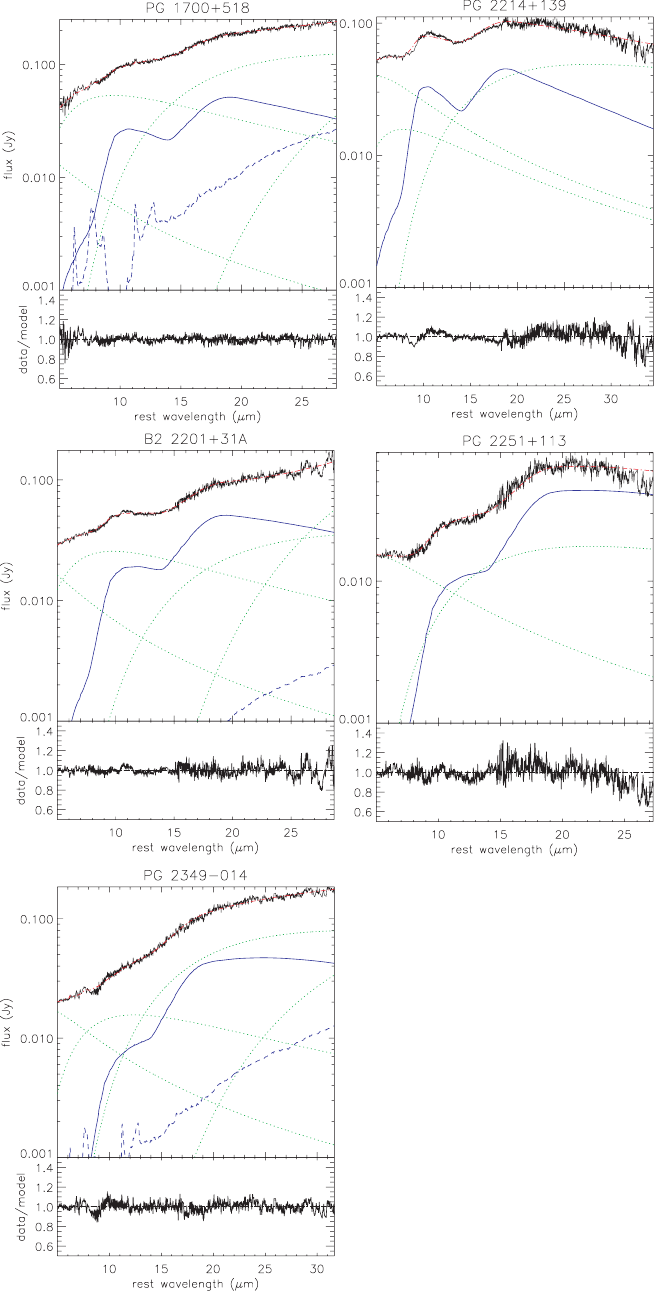}
\caption{continued}
\label{fig:fit results (D)}
\end{center}
\end{figure}

\clearpage
\begin{deluxetable}{lcccc}
\tabletypesize{\footnotesize}
\tablewidth{0pt}
\tablecaption{Results from NLR Model Fits \label{tab:resultsa}}
\tablehead{
\colhead{Object}&
\colhead{Best Fit NLR Model}&
\colhead{log($\frac{\mbox{Incident Flux}}{\mbox{\ergcms}}$)}&
\colhead{Cloud Distance}&
\colhead{Covering Factor\tablenotemark{1}}\\
\colhead{}&
\colhead{\#\tablenotemark{2}}&
\colhead{}&
\colhead{pc}&
\colhead{}\\
}
\startdata
PG0026+129          &6 &3.96  &54     & 0.25\\
PG0050+124 (IZw1)   &2 &5.16 &9      & 0.42\\ 
PG0838+770          &4 &4.56 &15     & 0.27\\
PG0953+414          &6 &3.96 &91     & 0.23\\
PG1001+054          &9 &2.96 &107    & 1.16\\
PG1004+130          &4 &4.56 &52     & 0.33\\
PG1116+215          &4 &4.56 &47     & 0.14\\
PG1126-041 (Mrk1298)&6 &3.96 &21     & 0.20\\
PG1229+204 (Mrk771) &5 &4.26 &21     & 0.14\\
PG1302-102          &6 &3.96 &98     & 0.41\\
PG1309+355          &6 &3.96 &64     & 0.50\\
PG1411+442          &4 &4.56 &18     & 0.17\\
PG1426+015          &5 &4.26 &30     & 0.21\\
PG1435-067          &3 &4.86 &14     & 0.17\\
PG1440+356 (Mrk478) &4 &4.56 &16     & 0.14\\
PG1613+658 (Mrk876) &5 &4.26 &40     & 0.24\\
PG1617+175          &3 &4.86 &13     & 0.12\\
PG1626+544          &2 &5.16 &11     & 0.10\\
PG1700+518          &5 &4.26 &124    & 0.15\\
PG2214+139 (Mrk304) &4 &4.56 &20     & 0.10\\
B2 2201+31A         &6 &3.96 &229    & 0.09\\
PG2251+113          &7 &3.56 &263    & 0.17\\
PG2349-014          &8 &3.26 &229    & 0.09\\
\enddata
\tablecomments{\\
(1) --- estimated covering factor assuming H$_0$=70km\,s$^{-1}$\,Mpc$^{-1}$, 
$\Omega_m=0.3$ and $\Omega_{\Lambda}=0.7$ \\
(2) -- the model number is a running number (1-13) identifying the NLR model. Larger model numbers correspond to cooler dust (lower incident fluxes). In table \ref{tab:NLRmodels} the model numbers are listed together with the respective ionization parameters and the assumed incident fluxes.\\
}
\end{deluxetable}

\begin{deluxetable}{lccr}
\tabletypesize{\footnotesize}
\tablewidth{0pt}
\tablecaption{Contributions to the total flux
\label{tab:resultsb}} 
\tablehead{
\colhead{Object}&
\colhead{NLR 6-25 $\mu$m}&
\colhead{NLR 15 $\mu$m}&
\colhead{M82 6-25 $\mu$m}\\
\colhead{}&
\colhead{$\%$}&
\colhead{$\%$}&
\colhead{$\%$}\\
\colhead{}&
\colhead{(1)}&
\colhead{(2)}&
\colhead{(3)}\\
}
\startdata
PG0026+129          &51 &63 &$<$5    \\
PG0050+124 (IZw1)   &19 &15 &5    \\  
PG0838+770          &21 &19 &11   \\   
PG0953+414          &44 &60 &$<$15    \\   
PG1001+054          &41 &46 &$<$8    \\   
PG1004+130          &44 &44 &$<$5    \\   
PG1116+215          &22 &28 &$<$9  \\   
PG1126-041 (Mrk1298)&15 &17 &7    \\   
PG1229+204 (Mrk771) &24 &25 &$<$4    \\    
PG1302-102          &29 &35 &$<$3    \\    
PG1309+355          &44 &50 &$<$13  \\    
PG1411+442          &16 &19 &4    \\   
PG1426+015          &23 &25 &5    \\   
PG1435-067          &29 &35 &$<$8    \\   
PG1440+356 (Mrk478) &16 &15 &20   \\   
PG1613+658 (Mrk876) &20 &24 &13   \\      
PG1617+175          &18 &17 &$<$5  \\    
PG1626+544          &37 &38 &$<$9    \\
PG1700+518          &19 &24 &$<$4 \\    
PG2214+139 (Mrk304) &28 &34 &$<$3   \\     
B2 2201+31A         &31 &47 &$<$3 \\       
PG2251+113          &40 &54 &$<$5    \\       
PG2349-014          &22 &26 &3       
\enddata
\tablecomments{\\
--- The total flux is measured across the same wavelength range as for the models.\\
(1) --- contribution of the NLR model to the total flux between 6 and 25 $\mu$m (rest-frame)\\
(2) --- contribution of the NLR model to the total flux integrated between 14.95 and 15.05 $\mu$m (rest-frame)\\
(3) --- contribution of the M82 template to the total flux between 6 and 25 $\mu$m (rest-frame). Upper limits have been estimated from a conversion of the (3$\sigma$) 7.7PAH upper limits presented in paper I. \\}
\end{deluxetable}

\begin{figure}
\includegraphics[width=400pt]{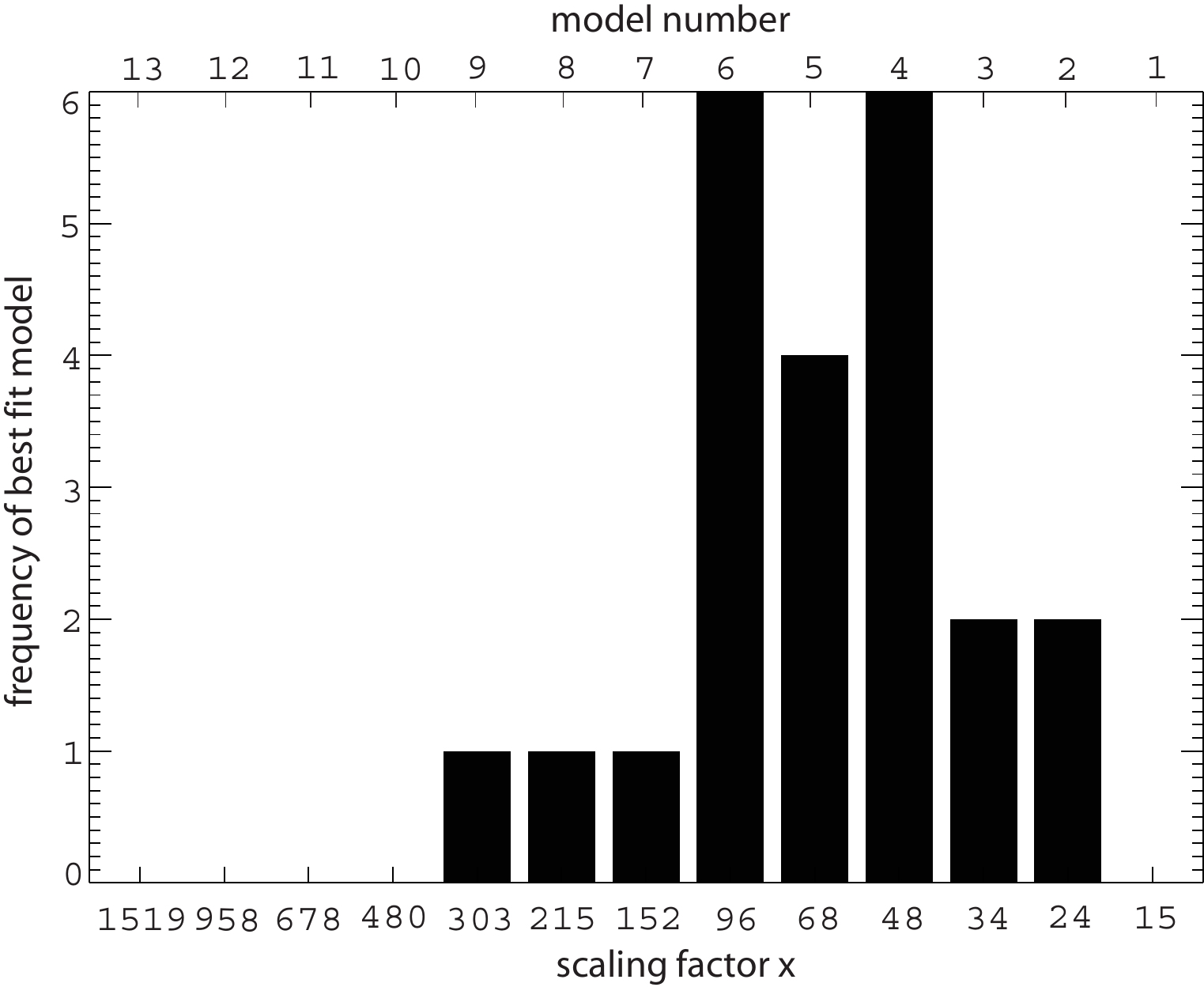}
\caption{Best fit NLR model distribution: Each model is related to a scaling law $R_{dust}\simeq x\cdot(L_{bol46})^{1/2}$pc for the dust cloud distance, where $x$ is the scaling factor and $L_{bol46}$ the bolometric AGN luminosity in units of 
$10^{46}$ $erg s^{-1}$.  }
\label{fig:statistics}
\end{figure}

\begin{figure}
\begin{center}
\includegraphics[angle=90,width=450pt]{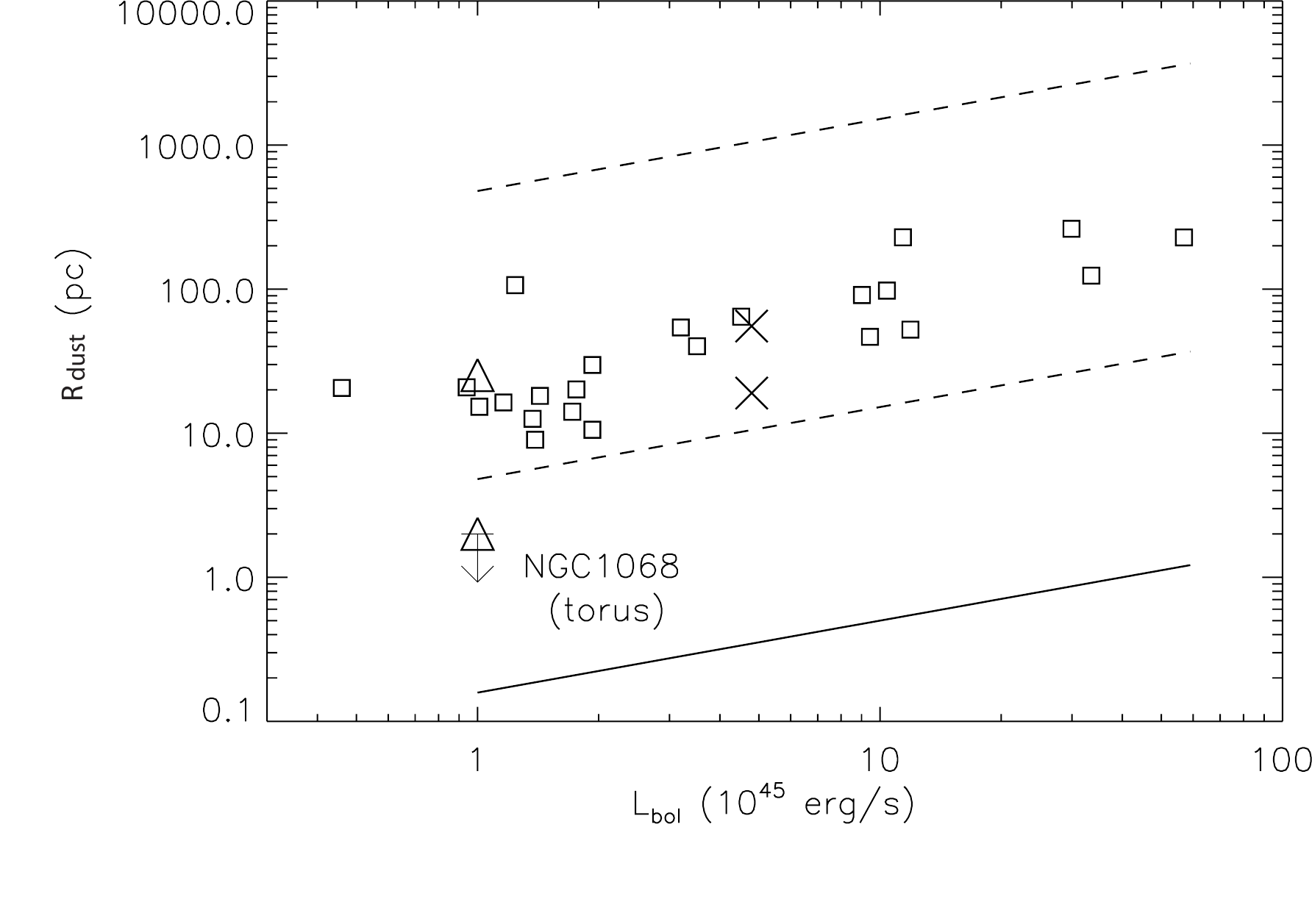}
\caption{NLR cloud distance versus $L_{bol}$,
compared to the dust sublimation distance $R_{sub}$ (solid line). 
On average, $R_{dust}$ is 170 times larger than $R_{sub}$. The dashed lines indicate the full distance range covered by the NLR models. The triangles indicate the upper limit for the torus size of NGC 1068 \citep{Jaffe04} (lower triangle) and the expected silicate dust distance at this AGN luminosity (upper triangle). The two crosses demonstrate the agreement for the silicate dust distance estimates for PG 0804+761 by \citet{marshall07} (lower cross) and based on our average scaling relation (upper cross) considering the general scatter of the estimated cloud distances.}
 \label{fig:distance}
\end{center}
\end{figure}

\begin{figure}
\includegraphics[angle=90,width=400pt]{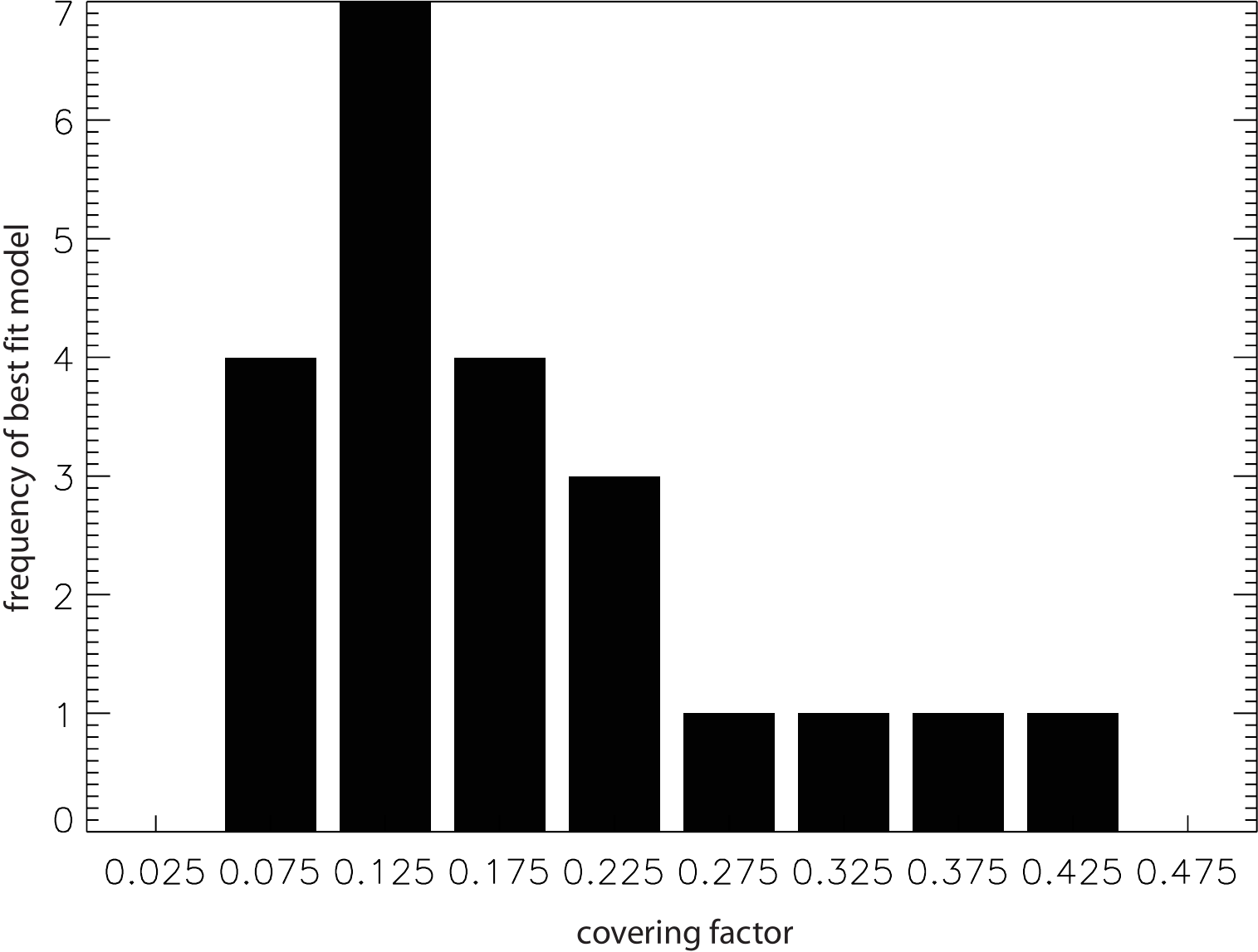}
\caption{Distribution of the derived covering factors (excluding PG1001+054, which has a derived covering factor of 1.16). Each bar has a width of 0.05 .  }
\label{fig:covering}
\end{figure}

\section{Discussion}
\label{sec:discussion}

We have successfully fitted the MIR \spitzer\ IRS spectra of 23 PG QSOs using cool NLR components to account for the silicate dust emission, in addition to components 
representing the underlying AGN continuum and the host galaxy's star formation. We have
investigated the robustness of the fit results by showing that the fit quality has a clear optimum as a function of NLR model (i.e. distance from the central source, Fig.~\ref{fig:bestfit}), and by probing for the effect of additional 
foreground extinction even for these optically unobscured Type-1 QSOs. Fitting our spectra with multiple NLR components simultaneously to allow for a dust cloud distribution, we confirm the typical cloud distances derived using one single NLR component but note that we cannot exclude smaller contributions from different components.

Our main result is that the silicate emission may arise in an extended region. 
The use of NLR models in conjunction with an underlying continuum is 
already suggestive of a model combining a central compact and hot dust continuum source with silicate-emitting clouds overlapping the NLR. Investigating in 
more detail the physical implications of the fit results can test this model as well as permit a discussion in the context of other scenarios of AGN dust emission, like more classical torus models or scenarios invoking disk winds
\citep{koenigl94,elvis00,Elitzur06}.

\subsection{Dust cloud distances}
\label{subsec:distance}
The dust cloud distances we derive are on average 170 times larger than the dust sublimation distance, which for our PG QSO sample reaches at maximum $\sim$1 pc for the most luminous sources. From the distribution of the best fitting NLR models, we estimate an average scaling relation of $R_{dust}\simeq 80(L_{bol46})^{1/2}$ pc for the cloud-source distance using the expression $L_{bol}= 7\times L(5100)$. The resulting cloud-source distances range from $\sim$10 pc to $\sim$260 pc.
The estimated distances and the good fit quality using our NLR models suggest that the NLR may host the silicate dust seen in emission. 
Figure \ref{fig:statistics} shows that the best-fit model distribution covers quite a narrow model range. This narrow range is also the reason for the quite small scatter if we plot cloud distance versus bolometric AGN luminosity (Eqn.\ref{eq:dist}), as seen in figure \ref{fig:distance}, and may be one of the most important findings. It reflects the fact that the range of models (and corresponding values of $F_{in}$) that fit the data best is a small sub-set of the total number of models.

Our results are in good agreement with the estimates of \citet{marshall07} for the QSO PG 0804+761. These authors used a similar decomposition technique and estimated a silicate dust cloud distance of $\sim$ 19pc, a factor $\sim$ 2.9 below the $\sim$55pc obtained from our average distance-luminosity scaling relation. As an example, in NGC1068, MIR VLT interferometry \citep{Jaffe04} finds compact mid-infrared emission components within r$\lesssim$2pc, which is much smaller than the expected distance of silicate emitting dust for an AGN of this luminosity (see figure \ref{fig:distance}), and also much smaller than the 
additional larger scale emission seen in standard mid-infrared images of this
AGN \citep{cameron93,bock00,tomono}.
This discrepancy offers additional support for the idea of a separation between the torus and the silicate emission region. Also, the tens of percent of extended mid-infrared 
emission suggested in this and other mid-infrared imaging studies of nearby
Seyferts \citep{radomski03,packham05} are in plausible agreement with the
fraction of 6-25$\mu$m emission ascribed to the NLR by our fits to 
PG QSOs (with somewhat higher luminosity).
We emphasize that our method does not assume any special geometry. Thus, while we are able to reach reasonable conclusions about the distance of the silicate dust, we cannot derive information about the angular distribution of the dust.

\subsection{Covering factors}

The median covering factor we derive for our QSO sample is 0.16 (excluding PG1001+054), with the overall distribution seen in figure \ref{fig:covering}. These values agree with other estimates based on narrow emission line imaging and equivalent line width measurements, which suggest NLR covering factors of $<30\%$ (e.g. \citealt{kriss92,Netzer93}). We note that in our case, the derived covering factors depend on the assumed column density of our models. For this reason, the covering factor is uncertain due to the uncertainty of the real column density. Fitting our sources with lower column densities than that assumed here ($N($H$)=10^{21.5}$\cmii) leads to too large covering factors. On the other hand, increasing the column will only result in a small reduction in covering factor, since most of the incoming flux is absorbed by the dust in a cloud of $N($H$)=10^{21.5}$\cmii.
The covering factor is basically determined by the emitted flux of the model. A larger column depth means higher dust absorption and stronger IR emission, which results in a lower covering factor. However, the relationship is not linear due to the change of the dust opacity with wavelength and the cooling of dust at greater depths. 
Source variability between the epochs of optical and our \spitzer\ observations may introduce another uncertainty. On average, this effect should cancel out, but it may explain the large covering factor (1.16) derived for PG1001+0054.
Differences between single and multiple NLR models indicate further uncertainties in the covering factor, since the contribution of each NLR component is somewhat lower in the multiple NLR model fit (the flux in multiple NLR model fits can be shared by similar components).

The distance to the center of the hottest dust is related to the AGN bolometric luminosity via our average scaling relation.
If the cloud size is assumed to be constant, then its covering factor would decrease if the cloud is placed further away from the central source. Another possibility is that rather than having discrete clouds, the entire ionization cone could be filled with dust. 
In this case, the covering factor does not correlate with distance.
To test if the covering factor is correlated with the NLR contribution to the total IR flux or the bolometric AGN luminosity, we tested for a correlation between the covering factor and the NLR contribution to the 15 $\mu$m continuum source flux as well as to the total AGN IR luminosity. No such correlation was found. 
The absence of a correlation may also reflect the uncertainty in our estimate of the covering factor

\subsection{Silicate emission and NLR properties}
\label{sec:lines}
The typical distance found here for the silicate emitting dust is some
170 times larger than the dust sublimation distance and hence much further
than assumed in canonical torus models. However, it is also smaller than ``typical''
dimensions assumed for the NLR. For example, \citet{bennert02} 
have derived a luminosity dependent NLR dimension of roughly
R(NLR)=2.1L([OIII])$_{42}^{0.5}$ kpc, where L([OIII])$_{42}$ is the [OIII]5007\AA\ line luminosity in units of $10^{42}$ $erg/s^{-1}$.
For the sources in our sample, this translates to a typical NLR dimension of 2 kpc
(the full range covered in our sample, using this relationship, is about 1-3 kpc).
Schmidt et al (2003) later investigated this relationship and found NLR dimensions that are
about of factor 2 smaller for sources similar to the ones in our sample.
These relationships were also discussed in \citet{netzer04}
and shown to be inconsistent with dimensions derived for the most luminous AGN. 
Regardless of the exact value, it seems that the dimensions found here for the silicate
emitting dust are 5-10 times smaller than the NLR size derived from the [OIII]5007\AA\ 
line luminosity.

We have also looked at the line emission expected from gas clouds situated at the
distance of the silicate emitting dust. Observed line fluxes will be published in a forthcoming paper (Veilleux et al. 2008, in preparation). Our continuum fitting 
procedure can only indicate the cloud distance. The line emission depends on the level of ionization of the gas (the ionization parameter) and hence on the gas density.
While a detailed study of the NLR properties in QUEST QSOs is beyond the scope of
the present work, we mention here several of the more important conclusions.
\begin{itemize}
\item
Assuming a density of about $10^4 cm^{-3}$, gives for the typical incident flux found
here, an ionization parameter of about $10^{-0.5}$. For this ionization parameter,
the high excitation lines of [NeV] at 14.3 and 24.3 $\mu$m can reach their observed
luminosities, given the covering fractions we derive. However, the observed [NeV]24/[NeV]14 line ratio 
(typically found to be $\sim$1 in our sample) suggests that this density is too high.
\item
For a lower density gas, of about $10^3 cm^{-3}$, we get the required [NeV] line ratio but 
the implied ionization parameter is so high that the [NeVI]7.6 $\mu$m line
is predicted to be much stronger than observed.
\item
In all cases of $N<10^{4.5} cm^{-3}$, the gas is too highly ionized to produce the strong observed lines
of [OIV] 26 $\mu$m and [NeIII] 15.6 $\mu$m. These lines can be reproduced for much higher
densities (lower ionization parameter) but, in this case, the [NeV] lines are predicted to be well
below their observed luminosities.
\end{itemize}
In short,there is no way to produce all the strong observed NLR lines from a single 
component situated at roughly the same distance as the silicate emitting dust.
None of these conclusions are very sensitive to the gas metallicity, the exact
shape of the ionizing continuum, the relative reddening of optical and IR
emission lines, or the exact NLR geometry.

Given this analysis we conclude that the region emitting the strong silicate features
cannot represent the entire NLR. It can perhaps be associated with gas
in the innermost NLR and the region emitting `coronal' lines from highly ionized species. 
In this scenario the extended portion of the NLR that produces most of the [OIII]5007\AA\ 
emission, and probably also the intermediate ionization IR lines like
[NeIII] 15.6 and [OIV]26 $\mu$m, is of much larger dimensions, perhaps by an order
of magnitude or so.
All this will be discussed in more detail in a forthcoming paper.

While detailed models designed to fit the IR spectrum of the QUEST QSOs are not yet available, some examples of the above line relationships can be found in the recent \citet{groves} study of the IR spectrum of dusty NLR clouds.

\subsection{Silicate emission from torus models}

Our finding of an extended silicate emitting region does not necessarily  
contradict torus models, since modified torus models 
(e.g. Nenkova et al. 2002 - Nenkova, M., Sirocky, M., Ivezic, Z., Elitzur, M., 2007, submitted to ApJ)
allow for weak or absent silicate 
emission within a certain parameter range (e.g. a certain clump distribution).
This is also true for some face-on torus models with smooth dust distributions \citep{dullemond05,fritz06} if certain parameters are assumed. A compact torus 
might be present and contribute strongly to the MIR continuum but less 
to the observed silicate emission. 

Many of the published torus models cited above predict some level of silicate
emission for nearly face-on Type-1 configurations, but often with a higher 
ratio of the 10 and 18$\mu$m silicate features than required by the apparently `cool' dust found  
in our \spitzer\ spectra. In-depth comparison to the \spitzer\ spectra is needed and
ongoing. To our knowledge, the first published result of this type is the fit
of \spitzer\/-observed silicate emission features in two Type-1 AGN to a torus 
model presented by \citet{fritz06}. These authors assume a flared disc with a  
continuous dust distribution, whose inner radius is defined by the sublimation radius of the 
dust, plus host galaxy emission (reprocessed stellar emission from dust in the host galaxy). They account for the different sublimation temperatures of silicate 
($T_{sub}$=1000 K) and graphite ($T_{sub}$=1500 K) dust grains which can 
result in a up to five times smaller minimum radius for the graphite grains. 
This difference results in a dust layer composed only of graphite grains, 
which attenuates the X-ray and UV emission from the central source and is 
taken into account when the authors compute the silicate minimum radius. 

Some of their fits (e.g., for PG1229+204 (Mrk771) and PG 2214+139 (Mrk304)) of Type-1 AGN, 
which are constrained only by broadband photometry, do not reproduce the silicate emission features observed 
with {\it Spitzer} (compare Fig.\ref{fig:fit results (B)}).
For PG1229+204, their torus model shows silicate in absorption rather than in emission.
The same is true for PG2214+139, which shows strong silicate emission features in its {\it Spitzer} spectrum.
These discrepancies suggest that broad band data alone are not able to constrain properly the MIR dust properties, since the wavelength sampling is too low. 
In contrast, their fits to two Type-1 AGN with MIR \spitzer\ spectra 
(PG0804+761 and PG1100+772 = 3C249.1) are satisfactory for both the silicate
features and the broad band SEDs. 

Two elements of the \cite{fritz06} models may contribute
to this successful fit. First, the separate treatment of graphite and silicate
sublimation moves the innermost silicate dust out in 
radius and behind some shielding graphite-only dust, resulting in
cooler temperatures. Results will sensitively depend on the exact values 
assumed for the sublimation temperatures. Second, the 
two successful fits to \spitzer\ spectra are characterized by a very steep
decrease of the total radial column density from the equatorial plane of
the flared disc to its edge. While the column N$_H$ is several 
10$^{23}$cm$^{-2}$ on the equator, it is about two orders of magnitude less
near the surface. Both factors combined imply a noticeable covering factor
by dust with a column of several 10$^{21}$cm$^{-2}$  at radii of order 1.2-36pc, with 
the silicate grains setting in only at $\sim$5pc. While this
still falls short by a fair factor from the radii $\sim$75pc implied by our 
scaling relation (for comparison we assume here the bolometric AGN luminosity used in \citealt{fritz06}), it suggests that part of the successful fit by this torus 
model may be due to a related concept: a moderate column of dust away from the 
equatorial plane of the torus and well outside the sublimation radius.
Their fits to the MIR \spitzer\ spectra of PG0804+761 and especially PG1100+772 
are somewhat worse compared to our average fit quality, but we note that 
our assumed underlying continuum (black bodies) is less physical compared to a proper radiative transfer calculation.

A key challenge for torus models of silicate emission will be to explain 
silicate emission detected in some clear Type-2 AGNs 
\citep{sturm2006,teplitz06}. 

The MIR wavelength range is very important for constraining model parameters, 
since the silicate emission features are very sensitive to dust temperatures, geometries,
and density distributions. Salient AGN properties linked to the circum-AGN region include the Type
1/2 ratio and the distribution of obscuring columns as determined from
X-rays \citep[e.g.][]{risaliti99} or the optical/near-infrared obscuration of 
the BLR \citep[e.g.][]{veilleux97,lutz02}. Models explaining silicate emission 
in terms of the torus alone need to fit those as well. A silicate emitting inner 
NLR of low optical depth, on the other hand, obviously has to be combined with
obscuring structures that are plausibly related to the hot continuum in 
our decompositions. Finally, future high spatial resolution
mid-infrared studies through 8m-class diffraction limited imaging, and
interferometry with improved sensitivity will be essential to remove
remaining ambiguities between the contributions of torus and NLR to the
silicate emission.

Finally, while the match between our models and the source spectra is in general good, we find
for some sources small deviations between the modeled and observed $\sim$10 $\mu$m silicate peak position.
Most clearly this can be seen in the fit of PG1626+554, where the observed silicate emission peaks somewhat redward from the modeled 9.7 $\mu$m silicate peak. This has already been reported earlier (e.g., \citealt{sturm05}) and may be explained by our physical assumptions, such as the grain size distribution and the exact chemistry of the silicate grains.

\section{Conclusions}
\label{sec:conclusions}

Scenarios based on the unified model suggest that silicate emission in AGN arises mainly from the 
illuminated faces of the clouds in the torus at temperatures near sublimation. 
However, detections of silicate emission in Type 2 QSOs,
and the estimated cool dust temperatures, argue for an origin in a more extended region.
To investigate this issue, we have presented the \spitzer-IRS spectra of 23 QSOs.
We have matched physically-based models to the mid-infrared spectra and found that the
silicate emission observed in these objects can be reproduced by emission from clouds, outside the central torus.
This extended silicate-emitting region is possibly associated with 
the innermost NLR region or the intermediate dusty region proposed by \citet{Netzer93}.

The dust cloud distances found here scale with the AGN luminosity as $R_{dust}\sim 80\cdot L_{bol46}^{0.5}$ pc, with the bolometric luminosity $L_{bol46} $ given in units of $10^{46}$ $erg s^{-1}$.
We have estimated the median distance of the dust cloud
responsible for the silicate emission to be 40 pc, while for individual
sources distances up to 260 pc are possible. The smallest cloud distance is 9 pc
for PG 0050+124. The calculated covering factors for the dust clouds have a 
median of 0.16 and are in agreement with an NLR origin of the silicate
emission. Our models do not exclude the possibility of a torus contribution to the observed silicate emission, but rather emphasize the good agreement and perhaps the necessity of a larger-scale contribution to this emission.
Finally, future high spatial resolution infrared observations and further crosschecks including the comparison to Type 1/2 ratios as well as distributions and individual values for the obscuring columns in X-rays and the near infrared are needed to resolve the remaining ambiguities between torus and NLR emission.

\begin{acknowledgements}

We thank Tod Boroson for kindly allowing us to use his optical data for PG QSOs.
We are grateful for comments by the referee.

Funding for this work has been provided by the Israel Science Foundation grant 232/03.
H. N. acknowledges a Humboldt Foundation prize and thanks the host institution, MPE Garching.

This research has made use of the NASA/IPAC Extragalactic Database (NED) which is operated by the Jet Propulsion
Laboratory, California Institute of Technology, under contract with the National Aeronautics and Space Administration. 

This work is based on observations carried out with the {\it Spitzer Space
Telescope}, which is operated by the Jet Propulsion Laboratory, California
Institute of Technology, under NASA contract 1407. Support for this work
was provided by NASA through contract 1263752 (S.V.) issued by
JPL/Caltech.

\end{acknowledgements}

\end{document}